\documentclass[journal]{IEEEtran}
\usepackage{amsmath,amsfonts}
\usepackage{algorithmic}
\usepackage{algorithm}
\usepackage{array}
\usepackage[caption=false,font=normalsize,labelfont=sf,textfont=sf]{subfig}
\usepackage{textcomp}
\usepackage{stfloats}
\usepackage{url}
\usepackage{verbatim}
\usepackage{graphicx}
\usepackage{cite}
\usepackage{bm}
\usepackage{mathrsfs}
\usepackage[table,xcdraw]{xcolor} % for revision; table
\usepackage{multirow} %table
\usepackage[normalem]{ulem} % underline in table
\useunder{\uline}{\ul}{}
\begin{document}

\title{Geometric Prior Based Deep Human Point Cloud Geometry Compression}
% Geometric/Shape/Category-specific prior based human point cloud geometry compression

\author{Xinju~Wu,
Pingping~Zhang,
Meng~Wang,~\IEEEmembership{Member,~IEEE,}
Peilin~Chen,
Shiqi~Wang,~\IEEEmembership{Senior Member,~IEEE,} 
and Sam~Kwong,~\IEEEmembership{Fellow,~IEEE}

\thanks{
Xinju Wu, Pingping Zhang, Meng Wang, Peilin Chen, and Shiqi Wang are with the Department of Computer Science, City University of Hong Kong, Hong Kong, China (e-mail: xinjuwu2-c@my.cityu.edu.hk; ppingyes@gmail.com; mwang98-c@my.cityu.edu.hk; plchen3-c@my.cityu.edu.hk; shiqwang@cityu.edu.hk).

Sam Kwong is with the Department of Computing and Decision Sciences, Lingnan University, Hong Kong, China (e-mail: samkwong@ln.edu.hk).
}

}

\maketitle

\begin{abstract}
The emergence of digital avatars has prompted an exponential increase in the demand for human point clouds with realistic and intricate details. The compression of such data becomes challenging due to massive amounts of data comprising millions of points.
Herein, we leverage the human geometric prior in the geometry redundancy removal of point clouds to greatly promote compression performance. More specifically, the prior provides topological constraints as geometry initialization, allowing adaptive adjustments with a compact parameter set that can be represented with only a few bits. Therefore, we propose representing high-resolution human point clouds as a combination of a geometric prior and structural deviations. The prior is first derived with an aligned point cloud. Subsequently, the difference in features is compressed into a compact latent code. The proposed framework can operate in a plug-and-play fashion with existing learning-based point cloud compression methods. Extensive experimental results show that our approach significantly improves the compression performance without deteriorating the quality, demonstrating its promise in serving a variety of applications.
\end{abstract}


\section{Introduction}

\IEEEPARstart{R}{ecent} years have witnessed unprecedented growth in the demand for extended reality (XR) and metaverse, where users can interact as digital avatars in collective virtual spaces.
Concurrently, 3D scanning devices such as scanners and LiDAR have become more affordable and accurate, enabling the efficient creation of a realistic digital twin of a physical human.
While meshes have been the prevalent representation for virtual humans, generating highly detailed and lifelike meshes demands substantial computing power.
An efficient and versatile alternative to representing humans is a point cloud, which allows for more accessible and accurate 3D scanning and modeling of human bodies and faces with intricate details.

\begin{figure}[!t]
        \centering
        \includegraphics[width=\linewidth]{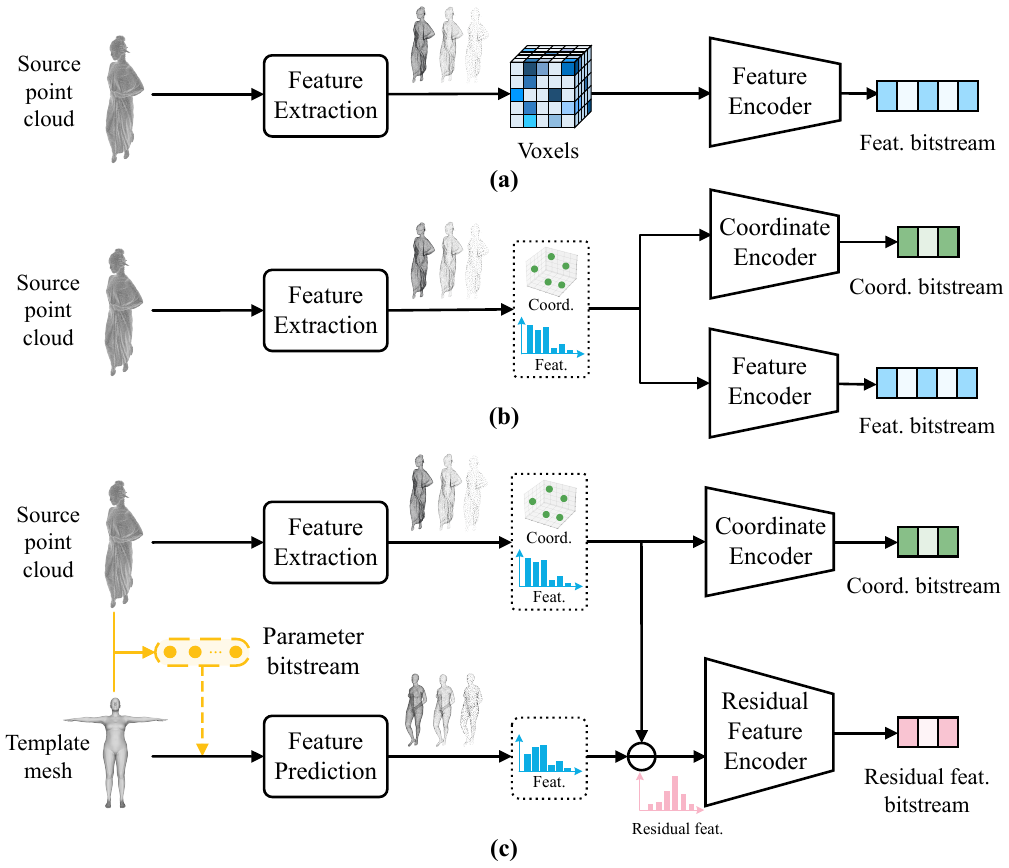}
        \caption{Comparisons of human point cloud geometry compression paradigms. Existing approaches directly compress the source point cloud and transmit (a) voxelwise features or (b) pointwise coordinates and features. (c) The proposed scheme incorporates a geometric prior to remove the redundancy at the feature level, followed by residual feature compression, yielding better compression performance.
        }
        \label{fig:structure}
\end{figure}

A point cloud is a collection of 3D data points that embody the surface geometry of an entity. Each point encompasses a coordinate in 3D space, along with additional information such as color, normal, and reflectance. To faithfully represent complex geometric shapes and structures, point clouds typically contain millions or billions of points. For instance, a high-resolution human point cloud from 8i's dataset~\cite{8iVFBv2} comprises $765,000$ points, each with a 30-bit coordinate $(x,y,z)$ and 24-bit color information $(r,g,b)$. This entails 11 MBytes of uncompressed storage for a point cloud and 3 GBytes for a 10-second point cloud video. The massive volume of data poses incredible challenges to the processing, transmission, and storage of high-quality point clouds. Therefore, it is imperative to develop point cloud compression (PCC) that can constrain data costs.

The traditional PCC methods developed by the Moving Picture Experts Group (MPEG)~\cite{graziosi2020overview,schwarz2019emerging} can be categorized into video-based PCC (V-PCC)~\cite{vpcc} for dynamic point clouds and geometry-based PCC (G-PCC)~\cite{gpcc} for static point clouds. V-PCC~\cite{vpcc} projects point clouds into two-dimensional (2D) planes and utilizes the hybrid video coding standard (e.g., High Efficiency Video Coding~\cite{sullivan2012overview}) for compression. G-PCC~\cite{gpcc} utilizes octree coding, trisoup coding, and predictive coding for geometry compression. On the other hand, deep learning based techniques have been successfully applied to PCC, leveraging the end-to-end training methodology~\cite{balle2017endtoend,balle2018variational}.
These approaches use an autoencoder architecture to encode a point cloud into a low-dimensional latent code. The latent code is then quantized, entropy-coded, and transmitted through a bitstream. The encoder has stacked downscaling blocks to reduce the number of points, while the decoder unfolds the latent code through upscaling blocks, reconstructing the original point sets.
The neural networks are trained toward the optimization of the rate-distortion (RD) performance, showing promising improvements for point cloud geometry compression.

\begin{figure}[!t]
    \centering
    \includegraphics[width=1\linewidth]{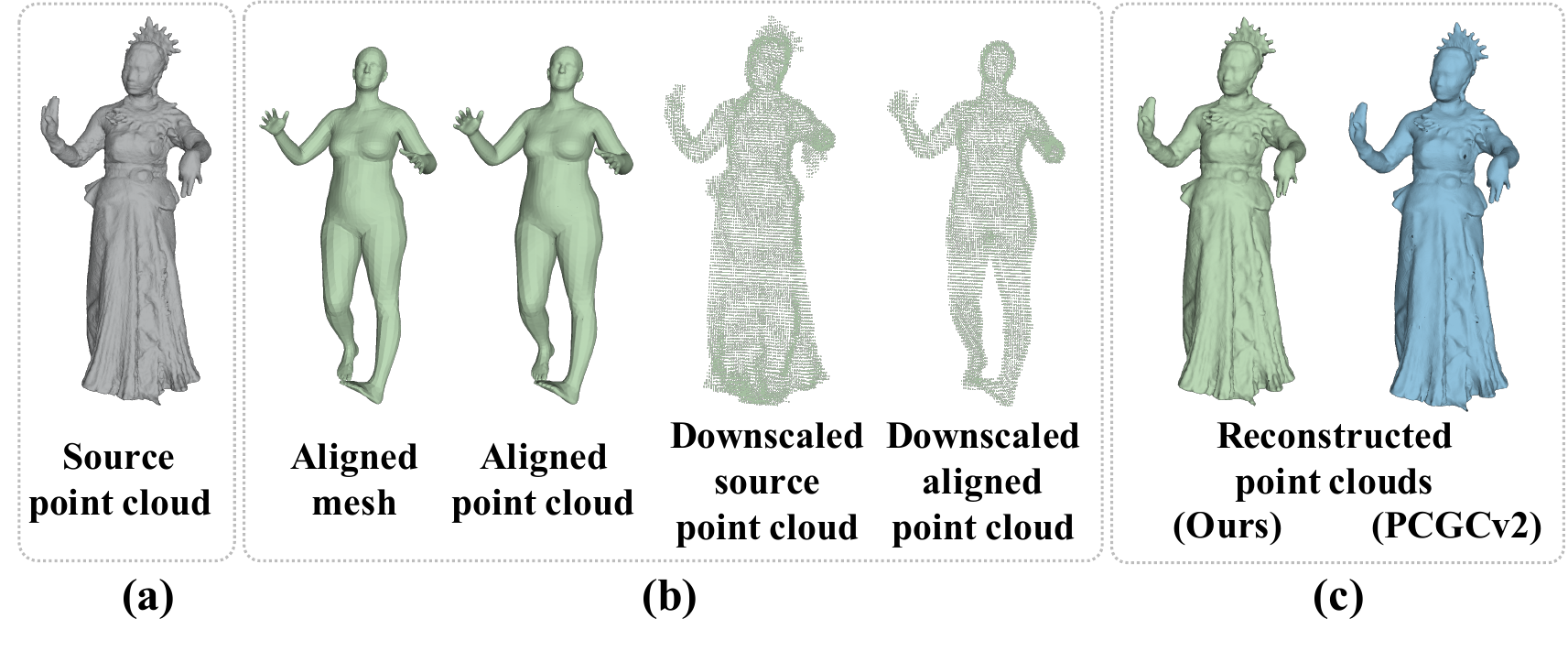}
    \caption{Visual quality comparisons of (a) a source point cloud, (b) intermediate 3D models generated by our approach, and (c) reconstructed point clouds at 0.125
    bits per point (bpp) for our approach and 0.152 bpp for PCGCv2~\cite{wang2021multiscale}.}
    \label{fig:visual_result}
\end{figure}

While prior works in learning-based PCC have shown promising results~\cite{quach2019learning,wang2021multiscale,wang2021sparse}, they fail to utilize essential geometric prior knowledge of 3D objects. As depicted in Fig.~\ref{fig:structure}(a) and \ref{fig:structure}(b), existing approaches encode the source point cloud by extracting inherent voxelwise features or pointwise coordinates and features without considering the underlying geometric structures in 3D shapes.
Human bodies exhibit well-defined components that can be effectively leveraged as explicit prior knowledge in compressing high-resolution human point clouds. Inspired by this, we propose a novel deep human point cloud geometry compression framework based on an explicit geometric prior, providing hard topological restrictions as an initialization of geometry, as illustrated in Fig.~\ref{fig:structure}(c).
Our framework leverages a compact set of geometric parameters encoded with only a few bits. These parameters control a shape prior model to generate an aligned point cloud, as depicted in Fig.~\ref{fig:visual_result}(b). By employing both source and aligned point clouds, our approach can effectively improve point cloud geometry compression over previous methods that directly compress the source point cloud.
Our main contributions are summarized as follows:
\begin{itemize}
    \item We propose a novel geometric prior based point cloud geometry compression framework in which human point clouds are compressed as the combination of a geometric prior and structure variations. Based on the prior, the redundancy is greatly removed at the feature level to improve the coding performance.
    \item We explore the 3D parametric model for PCC that realizes topological constraints as initialization for effective deep feature coding. This hybrid approach combines the strengths of mesh and point cloud representations, enabling high compression with the flexibility to represent complex shapes and fine-grained details.
    \item We incorporate our methodology in a plug-and-play manner for point cloud geometry compression. It is manifested that our approach yields superior RD performance compared to various baselines, exhibiting the superiority of our proposed scheme.
\end{itemize}

\section{Related Works}\label{sec:related}

\subsection{Traditional Point Cloud Geometry Compression}
The division of the point cloud based on an octree has been widely adopted in conventional approaches for compressing point cloud geometry, where only non-empty nodes among eight children continue to be subdivided. Mekuria~\textit{et al.}~\cite{mekuria2017design} first proposed a hybrid time-varying point cloud codec that serves as the anchor for MPEG PCC~\cite{graziosi2020overview,schwarz2019emerging}. In this codec, each intra-frame is progressively coded in the octree subdivision using 8-bit occupancy codes, while inter-frame redundancy is eliminated using the rigid transformation of 3D macroblocks in the octree voxel space. The MPEG has also developed the prevailing G-PCC and V-PCC standards~\cite{graziosi2020overview,schwarz2019emerging}. G-PCC~\cite{vpcc} relies on three techniques for geometry compression, including octree coding, trisoup coding, and predictive coding. Octree coding employs several modes to predict compact occupancy codes for isolated nodes, planes, and patterned regions, followed by an arithmetic coding engine. Trisoup coding aims to achieve lossy compression using a pruned octree for surface reconstruction and resampling. Predictive coding targets at large-scale LiDAR point clouds in low latency cases by pointwise prediction in tree-based traversal. In contrast, V-PCC~\cite{gpcc} adopts another line of compression by projecting 3D patches of point clouds onto the surfaces of a bounding box using 3D-to-2D projection, thus allowing for the reuse of existing video codecs~\cite{sullivan2012overview}. The projection result of the geometry component is 2D depth maps where each value represents the distance between each point and the projection plane.

Various techniques have been proposed to improve the geometry coding performance of both G-PCC and V-PCC. Specifically, for G-PCC~\cite{vpcc}, silhouette decomposition~\cite{ramalho2021silhouette}, dyadic decomposition~\cite{peixoto2020intraframe}, quad-tree and binary-tree partitions~\cite{zhang2020implicit}, and triangle construction~\cite{wang2021pointvoting} are used to enhance octree coding and trisoup coding. For V-PCC~\cite{gpcc}, block partitioning~\cite{ahmmed2021dynamica}, frame padding~\cite{li2021efficient}, and motion prediction~\cite{li2020advanced} are employed, along with rate-distortion optimization (RDO) based on geometric projection error~\cite{xiong2022efficient}. Additionally,  coding approaches detached from MPEG PCC~\cite{graziosi2020overview,schwarz2019emerging} have also been explored. For instance, Oliveira~\textit{et al.}~\cite{deoliveirarente2019graphbased} employed a graph-based transform for the enhancement layer and an octree-based approach for the base layer. Furthermore, Zhu~\textit{et al.} exploited region similarity~\cite{zhu2021lossy} and view-dependent projection~\cite{zhu2020viewdependent}, while Krivokuća~\textit{et al.}~\cite{krivokuca2020volumetric} introduced volumetric functions for geometry compression. In inter-frame compression, various methods for 3D motion compensation~\cite{dequeiroz2017motioncompensated,garcia2020geometry} and context-based arithmetic coding~\cite{thanou2016graphbased} have also been investigated.

\subsection{Learning-based Point Cloud Geometry Compression}

Recently, there has been a surge of interest in learning-based point cloud geometry compression. One direction involves the development of an efficient entropy model that leverages context, primarily for large-scale point clouds.
Huang~\textit{et al.}~\cite{huang2020octsqueeze} proposed a conditional entropy model with multiple ancestor nodes in the octree representation, whereas Que~\textit{et al.}~\cite{que2021voxelcontextnet} developed VoxelContext-Net, which utilizes information from neighboring octree nodes at the same depth level to improve local voxel context. Moreover, Fu~\textit{et al.}~\cite{fu2022octattention} utilized sibling nodes to expand context and an attention mechanism to emphasize key nodes, while children of sibling nodes and surface prior are further investigated in~\cite{fan2022multiscale}. For dynamic cases, Biswas~\textit{et al.}~\cite{biswas2021muscle} proposed an approach that models the probability of octree symbols and intensity values by exploiting spatial and temporal redundancy between successive LiDAR point clouds.

Another direction for learning-based point cloud geometry compression involves downsampling points in the encoder and recovering them in the decoder, extending end-to-end image~\cite{balle2017endtoend,balle2018variational} or video~\cite{hu2021fvc,lu2019dvc} compression techniques. Researchers have explored several methodologies in learning-based PCC, such as voxelization followed by 3D convolution, sparse convolution, and multilayer perceptron (MLP).
For example, Quach~\textit{et al.}~\cite{quach2019learning,quach2020improved} and Nguyen~\textit{et al.}~\cite{nguyen2021learningbased,nguyen2021multiscale} converted point clouds into 3D grids using voxelization and represented each voxel with an occupied or unoccupied state. Guarda~\textit{et al.} explored learning-based scalable coding for geometry~\cite{guarda2020deepa,guarda2020point} and obtained multiple RD points from a trained model using explicit quantization of the latent representation~\cite{guarda2020deep}. Milani~\cite{milani2021adae} introduced an adversarial autoencoding strategy to train the encoder. Wang~\textit{et al.}~\cite{wang2021lossy} proposed the PCGC framework, which includes preprocessing, autoencoder, and postprocessing modules, and used Voxception-ResNet (VRN)~\cite{brock2016generative} within the stacked unit and a hyperprior entropy model~\cite{balle2018variational}.
As a representative of sparse convolution based methods, a multiscale framework, PCGCv2, was proposed by Wang~\textit{et al.}~\cite{wang2021multiscale} based on sparse tensors to avoid the processing of massive empty voxels. To further improve the efficiency, they developed a more elaborate structure with downscaling and upscaling at each scale to calculate occupancy probability~\cite{wang2021sparse}, and this technique has been applied in LiDAR point clouds through neighborhood point attention~\cite{xue2022efficient}.
PointNet-based methods~\cite{liang2022transpcc,he2022densitypreserving, muzaddid2022variable,you2021patchbased} for point cloud compression employ set abstraction layers to extract local features, drawing inspiration from classification and segmentation tasks.
More specifically, self-attention layers in the transformer were first introduced by Liang~\textit{et al.}~\cite{liang2022transpcc}. Furthermore, density, local positions, and ancestor embeddings can be utilized to preserve local density information~\cite{he2022densitypreserving}.
Regarding inter-frame compression, Akhtar~\textit{et al.}~\cite{akhtar2022interframe} utilized sparse convolution to map the latent code of the previous frame to the coordinates of the current frame. Meanwhile, Fan~\textit{et al.}~\cite{fan2022ddpcc} proposed a multiscale motion flow fusion module for motion estimation and developed an adaptive weighted interpolation algorithm to further enhance motion estimation accuracy.

However, current learning-based point cloud geometry compression techniques typically neglect the prior knowledge of the source 3D model, resulting in geometric redundancy during the compression process. Despite that, incorporating prior knowledge into the source 3D model, such as its geometric properties, topology, or semantic information, can undoubtedly improve the coding efficiency.

\subsection{Representations from 3D priors}
Substantial attempts have also been made to retain explicit 3D geometric priors for 2D processing. Yang~\textit{et al.}~\cite{yang2023deep} manipulated a 3D morphable model as the face prior to transform a face between image space and UV texture space, which benefits image inpainting. Additionally, researchers have explored the enhancement of single-view images in the wild by concatenating regressed 3D shapes from 2D facial images and decoding results from face embedding~\cite{lin2020highfidelity}, or decomposing human and object images into 3D representations such as depth, normals, and albedo~\cite{wimbauer2022derendering,wu2020unsupervised}. In compression research, Chen~\textit{et al.}~\cite{chen2023interactive} recently proposed an interactive face video coding framework that converts inter frames into 3D meshes and projects them in the decoder, demonstrating promising performance in ultralow bitrate face communications.
Regarding image coding, segmentation maps and sketches~\cite{hoang2020image, 10032603} are relevant 2D external representations that can provide complementary shape cues for improving the coding performance.

For 3D processing, various methods have been developed to leverage 3D geometric prior information. Self-prior~\cite{hanocka2020point2mesh,wei2021deep} is utilized to model repeating geometric structures and leverage self-correlation across multiple scales with fine-grained details. In~\cite{smirnov2021learninga}, a parameterized 3D representation of Coons patches is used to represent a 3D object, which is optimized iteratively based on a deformable parameterized template model with a minimal number of control points. For human data, the skinned multi-person linear model (SMPL)~\cite{loper2015smpl} is an expressive 3D full-body template model that can be utilized as a 3D prior. In~\cite{xu2023occlusionaware}, the predicted parameters from the SMPL model are fed to a recognition module for improved pose estimation. Despite the increasing use of 3D priors in various applications, few attempts have been made to incorporate 3D priors in point cloud compression.

\section{Methodology}\label{sec:method}

\subsection{Overview}

\begin{figure*}[!t]
    \centering
    \includegraphics[width=\linewidth]{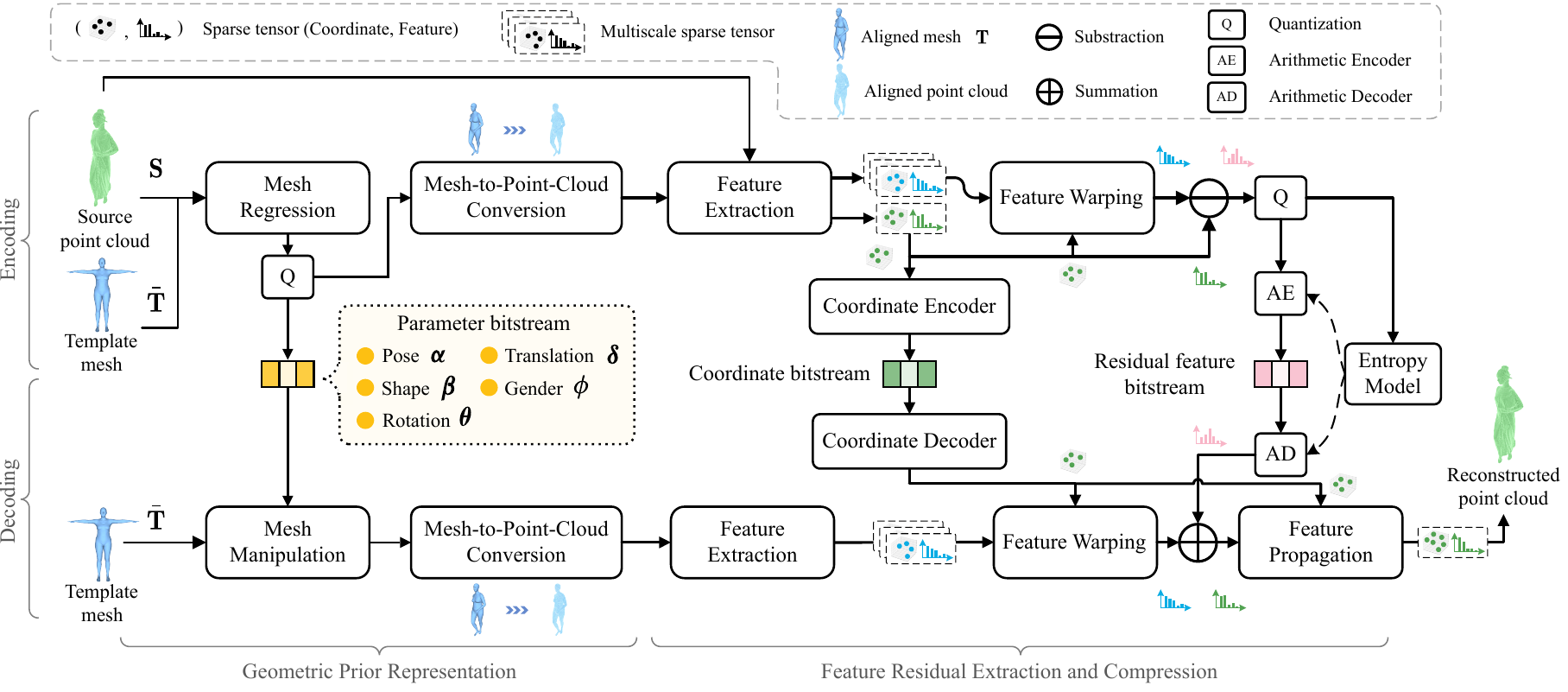}
    \caption{Overview of our proposed framework that involves a two-stage process for geometric prior representation and feature residual compression.
    Given a source point cloud $\mathbf{S}$, we first regress an aligned mesh $\mathbf{T}$ that can be driven by a set of parameters from a deformable template mesh $\bar{\mathbf{T}}$.
    During encoding, these parameters are further quantized into a compact bitstream, allowing for the manipulation of the template mesh's pose and shape during decoding.
    Regarding the next stage, we extract features from both the source point cloud and an aligned point cloud based on the sparse tensors that comprise coordinates and features. We then warp the features of the aligned point cloud onto the coordinates of the source point cloud, subsequently calculating residual features. These residual features are further encoded with guidance from an entropy model. The decoder, situated at the lower part of the framework, processes bitstreams to initiate the decoding process.
    }
    \label{fig:framework}
\end{figure*}

In this work, we develop a learning-based human point cloud geometry compression approach that leverages the human geometric priors to improve compression performance. The general architecture of the proposed scheme is shown in Fig.~\ref{fig:framework}. Specifically, the first stage of encoding involves fitting the source point cloud from a predefined template to derive a compact set of parameters representing the input geometry. Only the parameters need to be encoded and conveyed in the bitstream because the body modeling strategy and the mesh template are available during encoding and decoding.
However, there remain local geometric differences as the alignment of the source point cloud and the general template mesh focuses primarily on global shape and pose rather than perfect local correspondence.
To address this issue, we develop a second stage for feature residual extraction and compression. The goal is to encode local geometric variations through feature embeddings. To further reduce the size of the embeddings, we leverage the similarity between source and aligned point clouds.
Specifically, we perform feature warping operations before entropy encoding. This approach allows us to encode the residual features between corresponding aligned and source points rather than their full absolute features. Encoding residuals in the feature domain is more efficient as it captures remaining geometric variations without repeating shared details already established globally.
Thus, the coordinates of a downscaled source point cloud are losslessly encoded and transmitted.

Notably, the stage of feature residual extraction and compression has the ability to automatically accommodate input with different shapes without relying on human prior assumptions. Restricted memory available on GPU hardware makes it infeasible to feed entire high-resolution point clouds into the training process. Therefore, we partition both the source point cloud and the corresponding aligned point cloud into four equal blocks. These blocks serve as the input for the subsequent feature residual extraction and compression stage, which is trained end-to-end. During inference, we utilize a full unpartitioned point cloud as input.
Moreover, we adopt the same modules with the same weights for mesh-to-point-cloud conversion, feature extraction, and feature warping in the decoding process to maintain consistency on both sides, ensuring accurate reconstruction of high-quality point clouds.

\subsection{Geometric Prior Representation}

We employ the SMPL model~\cite{loper2015smpl} as our geometric prior, leveraging its compact and flexible representation to construct a comparable human point cloud that closely matches the shape and pose of the source point cloud.
The mean template used in our work can be manipulated by a collection of parameters,
\begin{equation}
        \bm{\Sigma} = \{\bm{\alpha}, \bm{\beta}, \bm{\theta}, \bm{\delta}, \phi \},
    \label{eq:param}
\end{equation}
where $\bm{\alpha} \in \mathbb{R}^{69}$, $\bm{\beta} \in \mathbb{R}^{10}$, $\bm{\theta} \in \mathbb{R}^{3}$, $\bm{\delta} \in \mathbb{R}^{3}$, and $\phi \in \mathbb{R}$ represent pose, shape, rotation, translation, and gender, respectively. The shape parameter determines regional variations, while the pose parameter controls joint rotations in the body.

The mesh manipulation module combines vertex deviations and surface deformation to model the human body, which enables extensively customizable and realistic representations.
With the predicted parameters, it is possible to represent vertex deviations from the template as
\begin{equation}
        \mathbf{V}=\bar{\mathbf{T}}+B_{\text{shape}}(\bm{\beta})+B_{\text{pose}}(\bm{\alpha}),
    \label{eq:offset}
\end{equation}
where $\bar{\mathbf{T}}$ denotes the mean template model. The functions $B_{\text{shape}}$ and $B_{\text{pose}}$ account for the effects of shape and pose deformations.
Based on proximity to the skeleton, surface deformation assigns weights to each vertex of the model,
\begin{equation}
    \mathbf{T}=H(\bm{\beta},\bm{\theta})\mathbf{V}+\bm{\delta},
    \label{eq:mesh_rec}
\end{equation}
where the function $H$ determines first the joint positions influenced by the shape and then the global rotation of these joints.

To regress a human point cloud with a parametric human model, we utilize the technique introduced by Zuo~\textit{et al.}~\cite{zuo2021self}.
The resulting predicted parameters are further quantized and encoded into a bitstream, as described in Section~\ref{sec:expt}-A.
To synchronize the encoder and decoder, we reconstruct an aligned mesh from quantized parameters during mesh manipulation. Subsequently, uniform sampling enables conversion of the aligned mesh into a corresponding aligned point cloud, as depicted in Fig.~\ref{fig:visual_result}(b).

\subsection{Feature Residual Extraction and Compression}
Using the aligned point cloud predicted from the geometric prior, we can promote the geometry compression performance by redundancy removal of the source point cloud. Specifically, we first extract low-dimensional feature embeddings of source and aligned point clouds separately. Because the aligned point cloud refers to a coarse approximation of the target positions, we perform warping operations within the feature space, a technique proven effective for deep video compression~\cite{hu2021fvc}. More precisely, we warp features of the aligned point cloud onto coordinates of the source point cloud using sparse convolution. This technique allows us to obtain compact residual features through feature subtraction, followed by the compression of residual features. Our proposed pipeline is versatile and can be applied in a plug-and-play manner by swapping out the feature extraction and warping modules with a variety of approaches. In our implementation, we inherit the feature extraction and warping techniques from the aforementioned deep point cloud compression approaches~\cite{wang2021lossy,wang2021multiscale,akhtar2022interframe} based on sparse convolution~\cite{SparseConvNet} to retain essential and critical point characteristics.

\begin{figure}[!t]
    \centering
    \includegraphics[width=\linewidth]{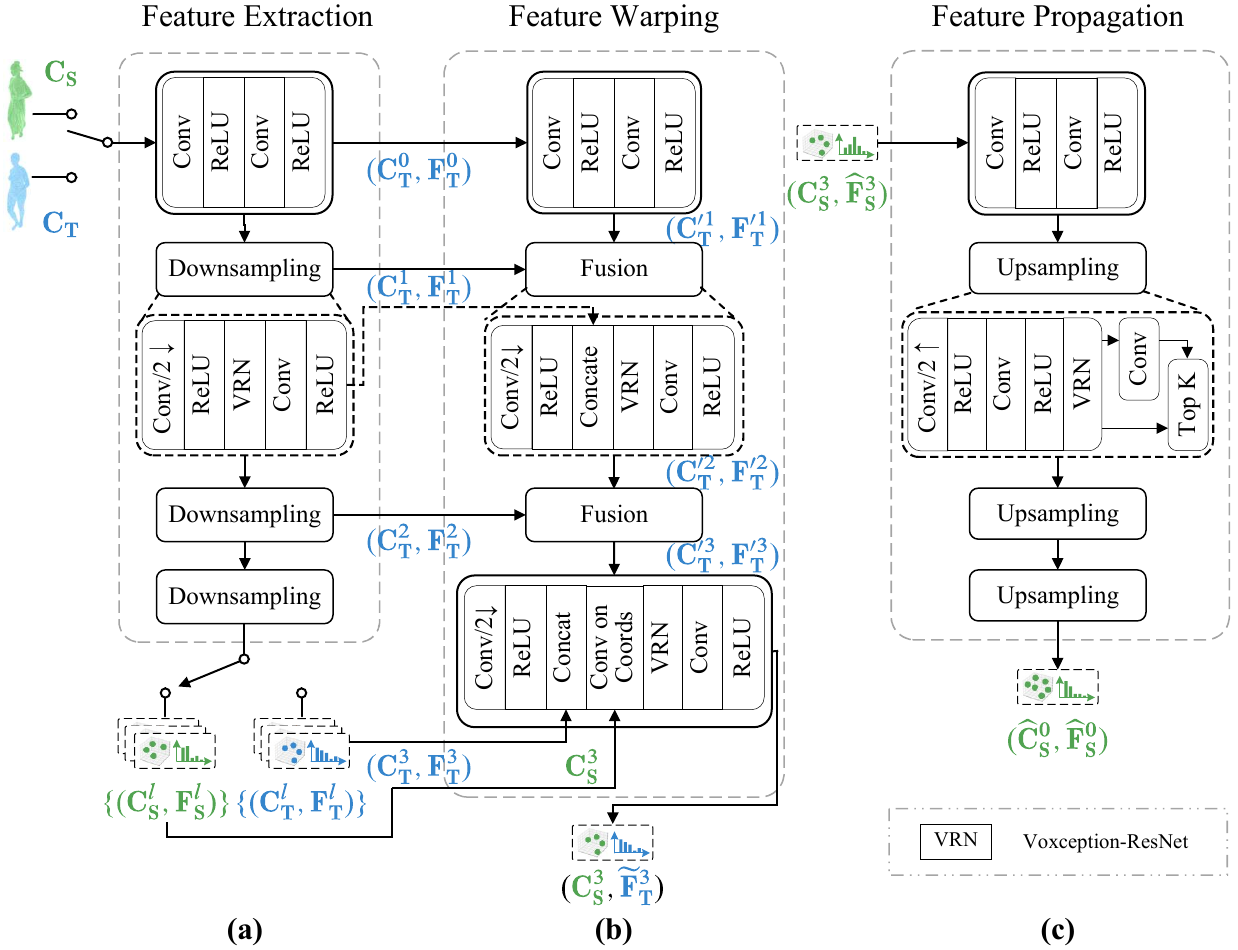}
    \caption{The network structure of (a) feature extraction, (b) feature warping, and (c) feature propagation modules. The input of the feature extraction module can be coordinates of the source point cloud $\mathbf{C}_\mathbf{S}$ or the aligned point cloud $\mathbf{C}_\mathbf{T}$. ``Conv/2$\downarrow$" and ``Conv/2$\uparrow$" represent the convolution and transposed convolution operations, respectively, with a stride of 2. ``Conv on Coords" convolves on target coordinates using a generalized transposed sparse convolution layer~\cite{SparseConvNet,akhtar2022interframe}. We consider an example with three scales, where $L=3$.}
    \label{fig:module}
\end{figure}

In sparse convolution techniques, intermediate outcomes between modules are represented by sparse tensors. Specifically, a sparse tensor $\mathcal{X}$ saves only non-zero elements using a coordinate-feature pair $\mathcal{X} \Leftrightarrow (\mathbf{C},\mathbf{F})$. Each non-zero coordinate $(x_i,y_i,z_i)\in \mathbf{C}$ corresponds to the associated feature $\mathbf{f}_i \in \mathbf{F}$. Sparse convolution in 3D space, formulated previously~\cite{SparseConvNet,choy20194d}, operates solely on non-zero input elements according to the expression,
\begin{equation}
    \mathbf{f}^\text{out}_\mathbf{u}=\sum_{\mathbf{i}\in\mathcal{N}^3(\mathbf{u}, \mathbf{C}^\text{in})} \bm{W}_\mathbf{i} \mathbf{f}^\text{in}_{\mathbf{u}+\mathbf{i}}\ \text{for} \ \mathbf{u} \in \mathbf{C}^\text{out},
    \label{eq:sparse_conv}
\end{equation}
where input and output coordinates $\mathbf{C}^\text{in}$ and $\mathbf{C}^\text{out}$ correspond to input and output feature vectors $\mathbf{F}^\text{in}$ and $\mathbf{F}^\text{out}$, respectively.
Let $\mathcal{N}^3(\mathbf{u}, \mathbf{C}^\text{in})$ denote the 3D kernel subset containing offset vectors from coordinate $\mathbf{u}$ to valid neighboring locations existing within the input $\mathbf{C}^\text{in}$, defined as $\mathcal{N}^3(\mathbf{u}, \mathbf{C}^\text{in})=\{\mathbf{i}|\mathbf{u}+\mathbf{i} \in \mathbf{C}^\text{in}, \mathbf{i}\in \mathcal{N}^3\}$. The kernel weight is denoted by $\bm{W}$.
In our implementation, the kernel is defined as a hypercube with a size of 3, namely $[-1,0,1]^3$, and Minkowski~\cite{choy20194d,gwak2020generative} is employed as the sparse inference engine.

\subsubsection{Feature Extraction} 
Using sparse convolution, our feature extraction module is designed to predict high-level embeddings in a bottom-up manner progressively. This process involves iteratively reducing the number of points while exponentially growing the receptive field size to aggregate information from wider areas.
As depicted in Fig.~\ref{fig:module}(a), our feature extractor contains successive downsampling blocks where each cascades a strided convolution, a VRN unit~\cite{brock2016generative,wang2021lossy}, and another convolution layer.
Specifically, the strided convolution reduces spatial resolution, while the subsequent convolution layer refines extracted features for optimal performance.
When situated between convolutions, the VRN unit~\cite{brock2016generative} leverages skip connections to reduce training information loss, along with parallel convolutions of diverse kernel sizes to provide network flexibility.
In this module, a point cloud is encoded into multiscale sparse tensors that contain coordinates and features.
After, we losslessly encode coordinates in the last scale using a coordinate encoder~\cite{gpcc}.

\subsubsection{Feature Warping}

\begin{figure}[!t]
    \centering
    \includegraphics[width=0.8\linewidth]{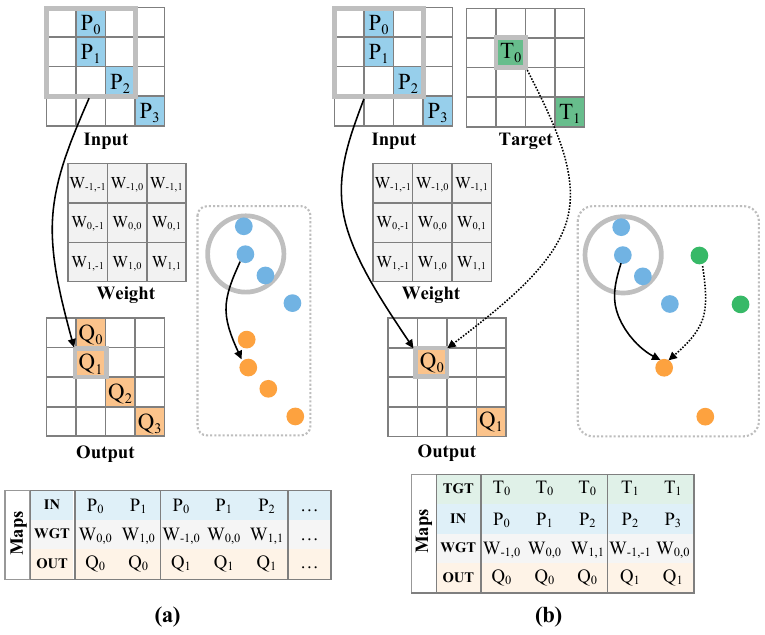}
    \caption{The 2D illustration of (a) vanilla sparse convolution and (a) the layer of convolution on coordinates.}
    \label{fig:conv}
\end{figure}

This module warps aligned point cloud features to source coordinates. As shown in Fig.~\ref{fig:module}(b), each scale initially concatenates primary sparse tensors $\mathcal{X}$ from the feature extractor module and auxiliary sparse tensors $\mathcal{X}'$ from the previous layer. This concatenation enhances downscaled outputs from the feature extraction module with informative points passed to subsequent blocks.
In the last scale, an additional ``convolution on coordinates'' layer warps aligned point cloud features onto downscaled source point cloud coordinates, enabling precise spatial alignment. We implement this process by a generalized sparse transposed convolution operation~\cite{SparseConvNet,akhtar2022interframe, gwak2020generative}.
As shown in Fig.~\ref{fig:conv}(a), vanilla sparse convolution operates only on the non-empty elements of the input sparse tensor. Our layer differs in that it takes two sparse tensor inputs, i.e., the input and the target sparse tensors, as depicted in Fig.~\ref{fig:conv}(b). The convolution is performed based on receptive fields centered on the target coordinates. This results in the output coordinates matching the target coordinates, rather than the original input coordinates.
The process can be succinctly represented as
\begin{equation}
    \mathcal{X}^L = G ( \mathcal{X}^L_\mathbf{T}, \mathbf{C}_{\mathbf{S}}^{L} ),
    \label{eq:conv_tr}
\end{equation}
where the function $G$ represents the convolving sparse tensor $\mathcal{X}^L_\mathbf{T}$ of aligned point cloud $\mathbf{T}$ on target coordinates $\mathbf{C}_\mathbf{S}^L$ of source point cloud $\mathbf{S}$ in the last scale $L$. The output sparse tensor is denoted as $\mathcal{X}^L \Leftrightarrow (\mathbf{C}_\mathbf{S}^L,\widetilde{\mathbf{F}}_\mathbf{T}^L)$, where $\widetilde{\mathbf{F}}_\mathbf{T}^L$ represents warped features of the aligned point cloud.
To further illustrate how this layer works, the bottom of Fig.~\ref{fig:conv}(b) shows example mappings for individual elements. For instance, for the point $Q_0$, its position remains the same as the target point $T_0$, while its features are computed from the weighted sum of receptive field centered on the input point $P_1$, which has the same position as the target point $T_0$.

\subsubsection{Residual Feature Calculation}
 
Having obtained warped features of the aligned point cloud, we can straightforwardly perform feature-level subtraction.
Recent works~\cite{akhtar2022interframe,fan2022ddpcc} on dynamic PCC have performed inter prediction within feature space. Following this vein, the proposed framework removes redundancy between source and aligned point clouds in the feature domain. Specifically, we subtract the warped features of the aligned point cloud $\widetilde{\mathbf{F}}_\mathbf{T}^L$ from the features of the source point cloud $\mathbf{F}_\mathbf{S}^L$, resulting in residual features. This process can be formulated as
\begin{equation}
    \Delta \mathbf{F}^L=\mathbf{F}_\mathbf{S}^L-\widetilde{\mathbf{F}}_\mathbf{T}^L,
    \label{eq:residual}
\end{equation}
where $\Delta \mathbf{F}^L$ denotes residual features in the scale $L$.
Feature-level alignment circumvents the challenge of directly determining offset correspondences in the Euclidean coordinate space between unmatched, unordered points. A similar attempt has also been proven effective for deep video compression~\cite{hu2021fvc}, reducing prediction errors caused by optical flow-based motion compensation in the pixel domain.

\subsubsection{Residual Feature Compression}

To compress the residual features, we apply vector quantization by adding a uniform quantizer, followed by entropy estimation of the residuals and arithmetic encoding for further compression. Specifically, additive uniform noise is incorporated for features during the training phase to enable approximation of the rounding operation while retaining differentiability for optimization purposes. During the inference phase, rounding is directly applied. After quantization, the entropy of the latent representation is estimated using an entropy bottleneck based on a non-parametric factorized model~\cite{balle2017endtoend}. The alternatives for the entropy model can be the hyperprior~\cite{balle2018variational} or joint autoregressive hierarchical priors~\cite{minnen2018joint}.

\subsection{Decoding}
The total bitstream comprises three components: geometric prior parameters, coordinates, and residual features. As illustrated in Fig.~\ref{fig:framework}, we reconstruct the original input geometry by generating an aligned point cloud, extracting features, and integrating residual features. Specifically, the parameters are first decoded to manipulate the template mesh available in both encoding and decoding. From the decoded parameters, an aligned mesh reconstructs and subsequently converts to an aligned point cloud. A feature extraction module then captures multiscale high-level embeddings from this aligned point cloud. We warp these extracted features onto the decoded downscaled coordinates of the source point cloud. Concurrently, we decode residual features from the bitstream and integrate them with the warped features to recover the source point cloud features. These features are subsequently propagated to upscale points approximating the source point cloud.

\begin{figure}[!t]
    \centering
    \includegraphics[width=1\linewidth]{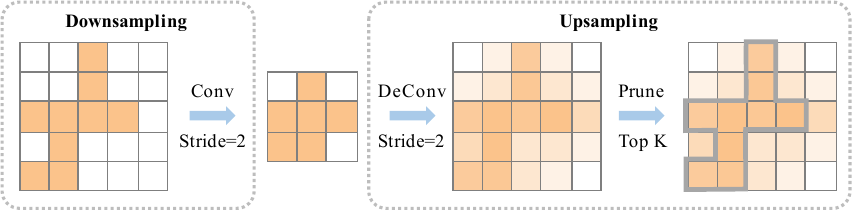}
    \caption{The decoder consists of repeated upsampling blocks with a transposed convolution layer that can generate more points than the input shape. A pruning layer is added in each upsampling block to probabilistically cull points with low predicted occupancy.}
    \label{fig:upsampling}
\end{figure}

The feature propagation module incorporates a transposed convolution layer with a stride of two in each upsampling block, as depicted in Fig.~\ref{fig:module}(c). This process allows for the upscaling of points while simultaneously preserving the sparsity pattern.
As illustrated in Fig.~\ref{fig:upsampling}, transposed convolutions may generate excess points, such that an extra convolution layer and a pruning layer are appended after VRN~\cite{brock2016generative}. The convolution layer computes occupancy probabilities, while the pruning layer removes points with low probability, retaining the top $K$ inputs. Here, $K$ equals the number of points in each scale.
Furthermore, we introduce hierarchical skip connections between the feature extraction and propagation modules during training. These connections provide multi-scale ground truth to learn efficient pruning in each upsampling block, effectively preserving information fidelity. The skip connections are removed during inference.

\subsection{Loss Function}
The objective of point cloud geometry compression is to minimize the number of needed bits while maintaining the maximum reconstruction quality of geometry. Toward this end, we optimize the RD tradeoff loss function,
\begin{equation}
    \mathcal{L}=\lambda R+D,
    \label{eq:RD_loss}
\end{equation}
where the Lagrange multiplier $\lambda$ balances rate $R$ and distortion $D$. Only the rate of feature residuals is represented by $R$, as the compression of both manipulation parameters and downscaled coordinates is excluded from the optimization process.
The point cloud reconstruction process can be formulated as a classification problem, where each point is classified as belonging to a 3D object or not~\cite{wang2021lossy}. The distortion between source and reconstructed point clouds is determined by the sum of widely used binary cross entropy (BCE)~\cite{gwak2020generative} in each scale,
\begin{equation}
    \begin{aligned}
        D &= \sum_{l=1}^{L}\text{BCE}^l(\mathcal{X} ^{l}_\mathbf{S},\widehat{\mathcal{X} }^{l}_\mathbf{S})\\
        &=\sum_{l=1}^{L} \frac{1}{N^l}\sum_{i=1}^{N^l}-b_i \log(p_i)-(1-b_i) \log(1-p_i),
    \end{aligned}
    \label{eq:distortion}
\end{equation}
where $\mathcal{X} ^{l}_\mathbf{S}$ and $\widehat{\mathcal{X} }^{l}_\mathbf{S}$ denote the source and decoded sparse tensors in scale $l=1,...,L$ with $L=3$ in our implementation, respectively, and $N^l$ represents the number of decoded points in scale $l$. A binary value $b_i$ indicates the occupancy of a decoded point $i$, and $p_i$ denotes the probability of that point being occupied.

\section{Experiments}\label{sec:expt}
\subsection{Implementation Details}
\subsubsection{Datasets} 
\begin{figure}[!t]
    \centering
    \includegraphics[width=\linewidth]{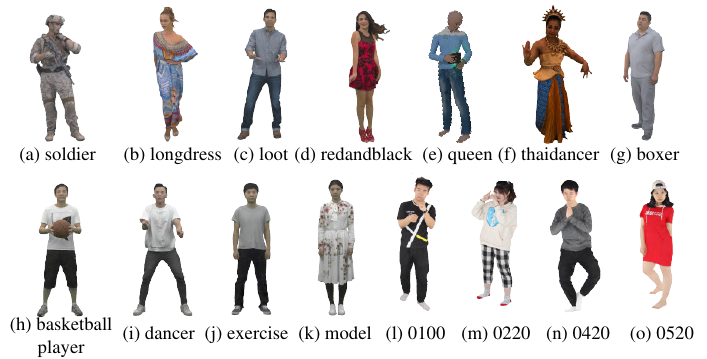}
    \caption{The training and testing datasets. Multiple frames of sequences from (a) \textit{soldier} to (e) \textit{queen} are utilized for training. A single frame of sequences from (f) \textit{thaidancer} to (o) 0520 is employed for testing. }
    \label{fig:datasets}
\end{figure}

We conduct a series of experiments on prevailing high-resolution human point cloud datasets, namely 8i Voxelized Full Bodies (8iVFBv2)~\cite{8iVFBv2}, Owlii dynamic human dataset (Owlii)~\cite{owlii}, THuman2.0~\cite{tao2021function4d}, and 8i Voxelized Surface Light Field (8iVSLF)~\cite{8ivslf}, as shown in Fig.~\ref{fig:datasets} and summarized in Table~\ref{tbl:datasets}. The 8iVFBv2 dataset~\cite{8iVFBv2} and the sequence \textit{queen} from MPEG PCC~\cite{ctc_vpcc, ctc_gpcc} are used for training. The former contains four sequences with 300 frames each, while the latter includes 250 frames.
As mentioned above, GPU memory constraints preclude training with full high-resolution point clouds. After the geometric prior representation, we partition the source into four patches at the midpoint along the x- and y-axes using a KD-tree, where the same partition boundaries are followed for the aligned point cloud.
During testing, we employ the entire human point clouds for inference, using point clouds from standardization committees: \textit{basketball player}, \textit{dancer}, \textit{exercise}, \textit{model} sequences from Owlii~\cite{owlii}, \textit{thaidancder} and \textit{boxer} from 8iVSLF~\cite{8ivslf}. To further demonstrate generalization, we also utilize high-quality human scans from the publicly available and challenging THuman2.0 dataset~\cite{tao2021function4d}. These scans are converted into point clouds with the midpoint subdivision algorithm and employed as testing data.
Table~\ref{tbl:datasets} provides details of the point clouds used for training and testing, where the last column denotes the number of bit values encoded along each axis of the 3D coordinate space.

\begin{table}[!t]
\caption{The detail of point clouds used in training and testing}
\centering
{
\renewcommand{\arraystretch}{1.05}
\footnotesize
\begin{tabular}{c|c||r|c|c}
\hline
\hline
Dataset                    & Point cloud                                                 & \# points & \# frames & \multicolumn{1}{c}{Precision} \\ \hline
\multirow{4}{*}{\begin{tabular}[c]{@{}c@{}}8iVFBv2\\\cite{8iVFBv2}\end{tabular}}   & soldier                                                     & 1,059,810 & 300       & 10                             \\ \cline{2-5} 
                           & longdress                                                   & 765,821   & 300       & 10                             \\ \cline{2-5} 
                           & loot                                                        & 784,142   & 300       & 10                             \\ \cline{2-5} 
                           & redandblack                                                 & 729,133   & 300       & 10                             \\ \hline
                           & queen                                                       & 1,006,509 & 250       & 10                             \\ \hline
\multirow{2}{*}{\begin{tabular}[c]{@{}c@{}}8iVSLF\\\cite{8ivslf}\end{tabular}}    & thaidancer                                                  & 979,857   & 1         & 10                             \\ \cline{2-5} 
                           & box                                                         & 994,546   & 1         & 10                             \\ \hline
\multirow{4}{*}{Owlii~\cite{owlii}}     & \begin{tabular}[c]{@{}c@{}}basketball\\ player\end{tabular} & 2,880,057   & 1         & 11                             \\ \cline{2-5} 
                           & dancer                                                      & 2,592,758   & 1         & 11                             \\ \cline{2-5} 
                           & exercise                                                    & 2,391,718   & 1         & 11                             \\ \cline{2-5} 
                           & model                                                       & 2,458,429   & 1         & 11                             \\ \hline
\multirow{4}{*}{\begin{tabular}[c]{@{}c@{}}THuman2.0\\\cite{tao2021function4d}\end{tabular}} & 0100                                                        & 2,391,718   & 1         & 10                             \\ \cline{2-5} 
                           & 0200                                                        & 847,940   & 1         & 10                             \\ \cline{2-5} 
                           & 0420                                                        & 766,152   & 1         & 10                             \\ \cline{2-5} 
                           & 0520                                                        & 770,210   & 1         & 10                             \\ \hline
\hline
\end{tabular}
}
\label{tbl:datasets}
\end{table}

\subsubsection{Performance evaluation}
The quantitative evaluation assesses the performance of our approach based on the RD criteria by computing Bjøntegaard delta rate (BD-Rate) and Bjøntegaard delta peak signal-to-noise ratio (BD-PSNR) results. The bitrate is calculated by total bitstreams of prior parameters, downscaled coordinates, and feature residuals, and the measurement is reported as bits per point (bpp). The geometric distortion is calculated by point-to-point (D1) and point-to-plane (D2) errors~\cite{ctc_vpcc, ctc_gpcc}. D1 computes the distance by connecting each point in a distorted point cloud and its closest points in the reference point cloud, and D2 derives a new distance vector by projecting the original distance vector along the normal direction. Following the MPEG common test conditions (CTC)~\cite{ctc_vpcc, ctc_gpcc}, we calculate the peak signal-to-noise ratio (PSNR) value over the symmetric D1 and D2. More specifically, we first apply the source point cloud as a reference to evaluate the decoded point cloud. Then, we swap them and compute the maximum PSNR value between these two paradigms to obtain the symmetric distortion.

\subsubsection{Training procedure}
The training procedure focuses on the coding of residual features produced by subtracting features of the source and aligned point clouds. We train seven models using different factors $\lambda$ in Eqn.~(\ref{eq:RD_loss}), specifically $\lambda \in\{0.2, 0.5, 1.1, 2.5, 6, 9, 13\}$.
The number of feature channels in the last layer of the encoder is set to 8.
Our methodology is accomplished on a machine with an NVIDIA GeForce RTX 3090 GPU in 24GB of memory, and we implement three scales in the hierarchical structure. We set the batch size as 8 and train the model for 64 epochs. The Adam optimizer is employed with weight decay, and the initial value is set to $10^{-4}$. Notably, compressing predicted geometric prior parameters and downsampled coordinates is not included in the training procedure.
Predicted parameters are obtained using the pretrained model from~\cite{zuo2021self} and then quantized to three decimal places before being written into a bitstream. For downsampled coordinates, we encode them losslessly using G-PCC~\cite{vpcc}.

\subsection{Performance Comparisons}
Here, we report the point cloud geometry coding performance and compare our proposed framework to other approaches to showcase the superiority of our methodology.

\subsubsection{Baselines}
To validate the effectiveness of our framework, we compare various point cloud geometry compression techniques, including traditional and learning-based approaches. G-PCC~\cite{vpcc} and V-PCC~\cite{gpcc} are representative techniques for conventional codecs, and PCGC~\cite{wang2021lossy} and PCGCv2~\cite{wang2021multiscale} are learning-based baselines.
Specifically, G-PCC and V-PCC are examined using the latest version available, i.e., TMC13v14 for G-PCC and TMC2v18 in all-intra mode for V-PCC. We compare two branches of G-PCC for geometry compression, namely, the octree-based and surface reconstruction-based (trisoup) schemes. Our quantization parameter settings for G-PCC (octree), G-PCC (trisoup), and V-PCC follow CTC~\cite{ctc_vpcc, ctc_gpcc}, with the bitstream compositions for attribute disregarded. For learning-based baselines, PCGC employs point cloud voxelization and stacked 3D convolutions to capture compact features, while PCGCv2 leverages sparse convolution layers in a multiscale manner. Moreover, our framework is versatile and compatible with a plug-and-play setup, and the feature extraction component shown in our framework utilizes the same network structure as PCGCv2. For fair comparisons, a factorized prior model~\cite{balle2017endtoend} is employed as the entropy model in the learning-based baselines and our approach.

\subsubsection{Experimental results}

\begin{table*}[!t]
    \caption{BD-Rate and BD-PSNR results against the baselines G-PCC (octree)~\cite{vpcc}, G-PCC (trisoup)~\cite{vpcc}, V-PCC~\cite{gpcc}, PCGC~\cite{wang2021lossy}, PCGCv2~\cite{wang2021multiscale} on datasets Owlii~\cite{owlii}, 8iVSLF~\cite{8ivslf}, and THuman2.0~\cite{tao2021function4d} using D1 and D2 errors~\cite{ctc_vpcc, ctc_gpcc}}
    \centering
    %\resizebox{\textwidth}{!}
    { % \linewidth
\footnotesize
    \begin{tabular}{cc||ccccc||ccccc}
    \hline
\hline
\multicolumn{1}{c|}{\multirow{2}{*}{Dataset}}   & \multirow{2}{*}{Sequence}                                   & \multicolumn{5}{c||}{BD-Rate with D1 PSNR (\%)}                                                                                                                             & \multicolumn{5}{c}{BD-PSNR with D1 (dB)}                                                                                                                             \\ \cline{3-12} 
\multicolumn{1}{c|}{}                           &                                                             & \begin{tabular}[c]{@{}c@{}}G-PCC\\(octree)\end{tabular} & \begin{tabular}[c]{@{}c@{}}G-PCC\\(trisoup)\end{tabular} & V-PCC           & PCGC            & PCGCv2          & \begin{tabular}[c]{@{}c@{}}G-PCC\\(octree)\end{tabular} & \begin{tabular}[c]{@{}c@{}}G-PCC\\(trisoup)\end{tabular} & V-PCC         & PCGC          & PCGCv2        \\ \hline
\multicolumn{1}{c|}{\multirow{2}{*}{8iVSLF}}    & boxer                                                       & -93.92                                                   & -93.62                                                    & -50.02          & -67.96          & -32.60          & 12.77                                                    & 7.89                                                      & 2.65          & 5.16          & 1.48          \\
\multicolumn{1}{c|}{}                           & thaidancder                                                 & -91.85                                                   & -87.86                                                    & -48.48          & -61.25          & -22.42          & 11.56                                                    & 7.81                                                      & 2.65          & 4.83          & 1.04          \\ \hline
\multicolumn{1}{c|}{\multirow{4}{*}{Owlii}}     & basketball player & -95.42                                                   & -98.36                                                    & -93.54          & -69.61          & -29.31          & 13.60                                                    & 9.03                                                      & 8.47          & 5.08          & 1.19          \\
\multicolumn{1}{c|}{}                           & dancer                                                      & -95.20                                                   & -97.75                                                    & -94.03          & -68.98          & -30.30          & 13.51                                                    & 9.19                                                      & 8.82          & 4.95          & 1.23          \\
\multicolumn{1}{c|}{}                           & exercise                                                    & -95.07                                                   & -98.22                                                    & -93.12          & -68.19          & -30.84          & 13.55                                                    & 8.81                                                      & 8.54          & 4.89          & 1.29          \\
\multicolumn{1}{c|}{}                           & model                                                       & -93.94                                                   & -94.10                                                    & -91.24          & -75.77          & -32.73          & 12.86                                                    & 8.70                                                      & 8.62          & 6.58          & 1.54          \\ \hline
\multicolumn{1}{c|}{\multirow{4}{*}{THuman2.0}} & 0100                                                        & -89.54                                                   & -75.01                                                    & -19.08          & -64.87          & -26.80          & 9.13                                                     & 5.02                                                      & 0.55          & 4.03          & 1.10          \\
\multicolumn{1}{c|}{}                           & 0220                                                        & -88.64                                                   & -75.72                                                    & -37.71          & -57.37          & -20.60          & 9.04                                                     & 5.15                                                      & 1.38          & 3.47          & 0.91          \\
\multicolumn{1}{c|}{}                           & 0420                                                        & -90.20                                                   & -77.84                                                    & -30.34          & -56.38          & -33.06          & 9.62                                                     & 5.22                                                      & 1.04          & 3.17          & 1.37          \\
\multicolumn{1}{c|}{}                           & 0520                                                        & -89.60                                                   & -74.59                                                    & -36.18          & -59.39          & -28.35          & 9.29                                                     & 5.06                                                      & 1.29          & 3.40          & 1.29          \\ \hline
\multicolumn{2}{c||}{\textbf{Average with D1}}                                                                 & \textbf{-92.34}                                          & \textbf{-87.31}                                           & \textbf{-59.37} & \textbf{-64.98} & \textbf{-28.70} & \textbf{11.49}                                           & \textbf{7.19}                                             & \textbf{4.40} & \textbf{4.56} & \textbf{1.25} \\

\hline
%\multicolumn{2}{c||}{}        &      &          & &  &  &                  &                                      &  & &        \\ 
\hline

\multicolumn{1}{c|}{\multirow{2}{*}{Dataset}}   & \multirow{2}{*}{Sequence}                                   & \multicolumn{5}{c||}{BD-Rate with D2 PSNR (\%)}                                                                                                                                                                                                    & \multicolumn{5}{c}{BD-PSNR with D2 (dB)}                                                                                                                                                                                                         \\ \cline{3-12} 
\multicolumn{1}{c|}{}                           &                                                             & \multicolumn{1}{c}{\begin{tabular}[c]{@{}c@{}}G-PCC\\(octree)\end{tabular}} & \multicolumn{1}{c}{\begin{tabular}[c]{@{}c@{}}G-PCC\\(trisoup)\end{tabular}} & \multicolumn{1}{c}{V-PCC} & \multicolumn{1}{c}{PCGC} & PCGCv2 & \multicolumn{1}{c}{\begin{tabular}[c]{@{}c@{}}G-PCC\\(octree)\end{tabular}} & \multicolumn{1}{c}{\begin{tabular}[c]{@{}c@{}}G-PCC\\(trisoup)\end{tabular}} & \multicolumn{1}{c}{V-PCC} & \multicolumn{1}{c}{PCGC} & \multicolumn{1}{c}{PCGCv2} \\ \hline
\multicolumn{1}{c|}{\multirow{2}{*}{8iVSLF}}    & boxer                                                       & -90.90                                                                       & -92.23                                                                        & -49.28                    & -64.72                   & -30.16                      & 12.08                                                                        & 8.99                                                                          & 3.01                      & 5.00                     & 1.59                       \\
\multicolumn{1}{c|}{}                           & thaidancder                                                 & -88.36                                                                       & -88.38                                                                        & -51.75                    & -61.64                   & -22.50                      & 10.87                                                                        & 8.43                                                                          & 3.20                      & 4.76                     & 1.17                       \\ \hline
\multicolumn{1}{c|}{\multirow{4}{*}{Owlii}}     & basketball player& -92.38                                                                       & -96.73                                                                        & -86.81                    & -70.48                   & -25.70                      & 12.46                                                                        & 9.18                                                                          & 8.64                      & 5.79                     & 1.27                       \\
\multicolumn{1}{c|}{}                           & dancer                                                      & -92.18                                                                       & -95.61                                                                        & -87.66                    & -70.25                   & -26.06                      & 12.35                                                                        & 9.23                                                                          & 8.96                      & 5.37                     & 1.27                       \\
\multicolumn{1}{c|}{}                           & exercise                                                    & -91.88                                                                       & -96.02                                                                        & -86.48                    & -68.13                   & -26.79                      & 12.36                                                                        & 9.30                                                                          & 8.71                      & 5.42                     & 1.35                       \\
\multicolumn{1}{c|}{}                           & model                                                       & -89.79                                                                       & -90.97                                                                        & -85.89                    & -68.29                   & -27.97                      & 11.27                                                                        & 9.02                                                                          & 8.85                      & 6.11                     & 1.50                       \\ \hline
\multicolumn{1}{c|}{\multirow{4}{*}{THuman2.0}} & 0100                                                        & -86.58                                                                       & -80.36                                                                        & -37.43                    & -72.03                   & -21.94                      & 9.32                                                                         & 6.38                                                                          & 1.60                      & 2.65                     & 0.98                       \\
\multicolumn{1}{c|}{}                           & 0220                                                        & -86.56                                                                       & -83.43                                                                        & -56.99                    & -58.65                   & -13.93                      & 9.47                                                                         & 7.16                                                                          & 3.23                      & 3.24                     & 0.64                       \\
\multicolumn{1}{c|}{}                           & 0420                                                        & -87.73                                                                       & -82.52                                                                        & -45.90                    & -66.95                   & -15.72                      & 9.96                                                                         & 7.08                                                                          & 2.27                      & 2.60                     & 0.74                       \\
\multicolumn{1}{c|}{}                           & 0520                                                        & -88.16                                                                       & -84.53                                                                        & -55.13                    & -71.29                   & -15.21                      & 9.85                                                                         & 7.21                                                                          & 2.89                      & 2.56                     & 0.68                       \\ \hline
\multicolumn{2}{c||}{\textbf{Average with D2}}                                                                 & \textbf{-89.45}                                                              & \textbf{-89.08}                                                               & \textbf{-64.33}           & \textbf{-67.24}          & \textbf{-22.60}             & \textbf{11.00}                                                               & \textbf{8.20}                                                                 & \textbf{5.13}             & \textbf{4.35}            & \textbf{1.12}              \\ 

\hline
\hline
        \end{tabular}

    }
    \label{tbl:RD}
\end{table*}

In Table~\ref{tbl:RD}, we report the BD-Rate and BD-PSNR results of the proposed framework against G-PCC (octree), G-PCC (trisoup), V-PCC, PCGC, and PCGCv2 with D1 and D2 errors as distortion and bpp as bitrate. Our approach achieves significant bitrate savings and BD-PSNR gains compared to these traditional and learning-based methods on human point clouds from various datasets. Specifically, our method outperforms G-PCC (octree) with an average of 92.34\% and 89.45\% bitrate savings in terms of D1 and D2, respectively. Significant improvement has also been noticed against G-PCC (trisoup) and V-PCC with more than 87\% and 59\% BD-Rate gains, respectively, regarding both distortion errors. Compared with learning-based methods such as PCGC and PCGCv2, we achieve 64.98\% and 28.70\% bitrate savings in terms of D1, respectively. In particular, our approach has approximately 1.26~dB gains over PCGCv2 on the 8iVSLF dataset, 1.31~dB on Owlii, and 1.17~dB on THuman2.0. As PCGCv2 shares the same feature extraction network structure as ours, the performance improvement is a clear indication of the effectiveness of incorporating geometric priors and residual features. Our approach also outperforms learning-based baselines with respect to D2 errors. 

\begin{figure*}[!t]
    \centering
    %0.45, 0.295
    \begin{minipage}[]{0.31\linewidth}
        \includegraphics[width=\linewidth]{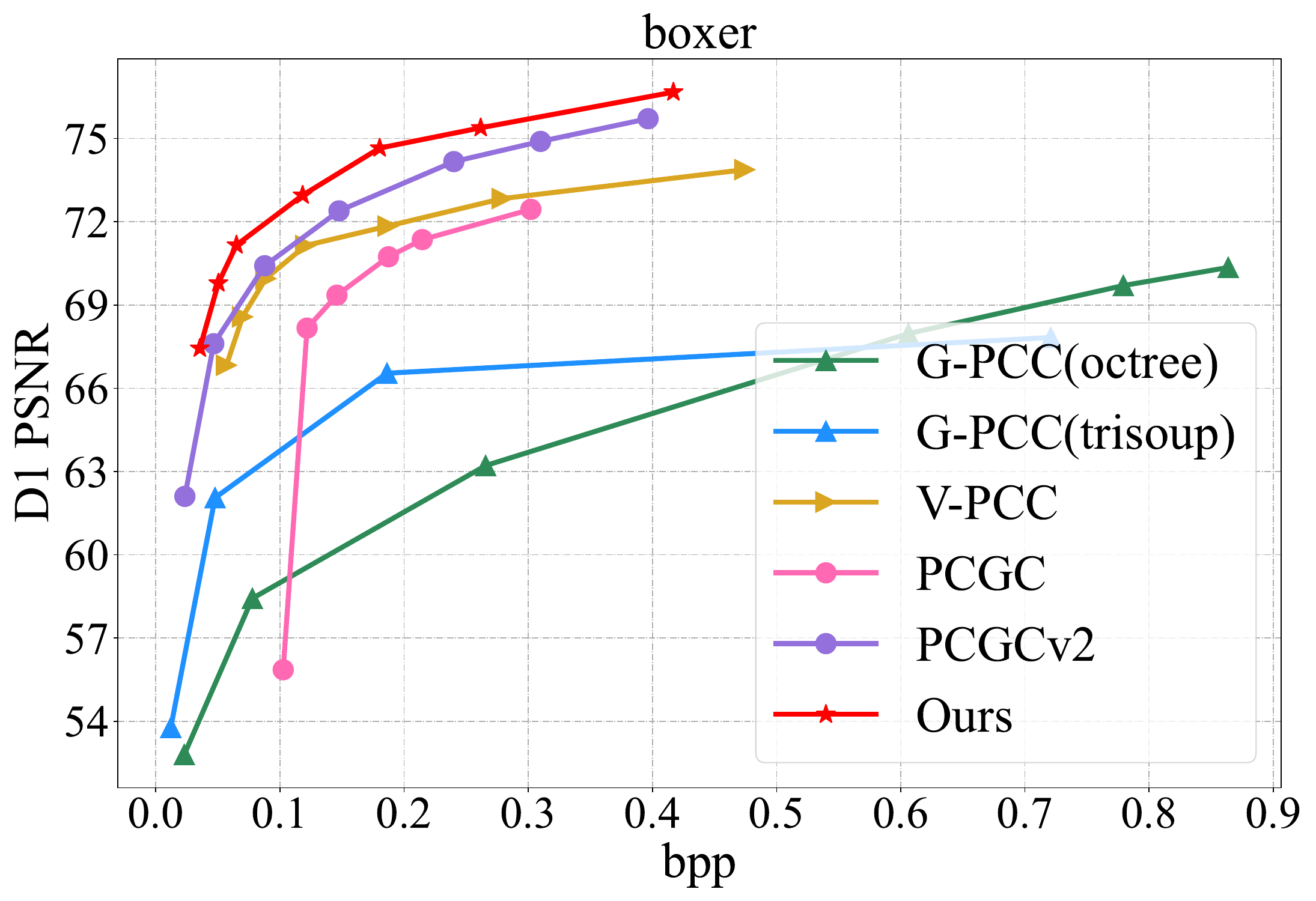}
    \end{minipage}
    \begin{minipage}[]{0.31\linewidth}
        \includegraphics[width=\linewidth]{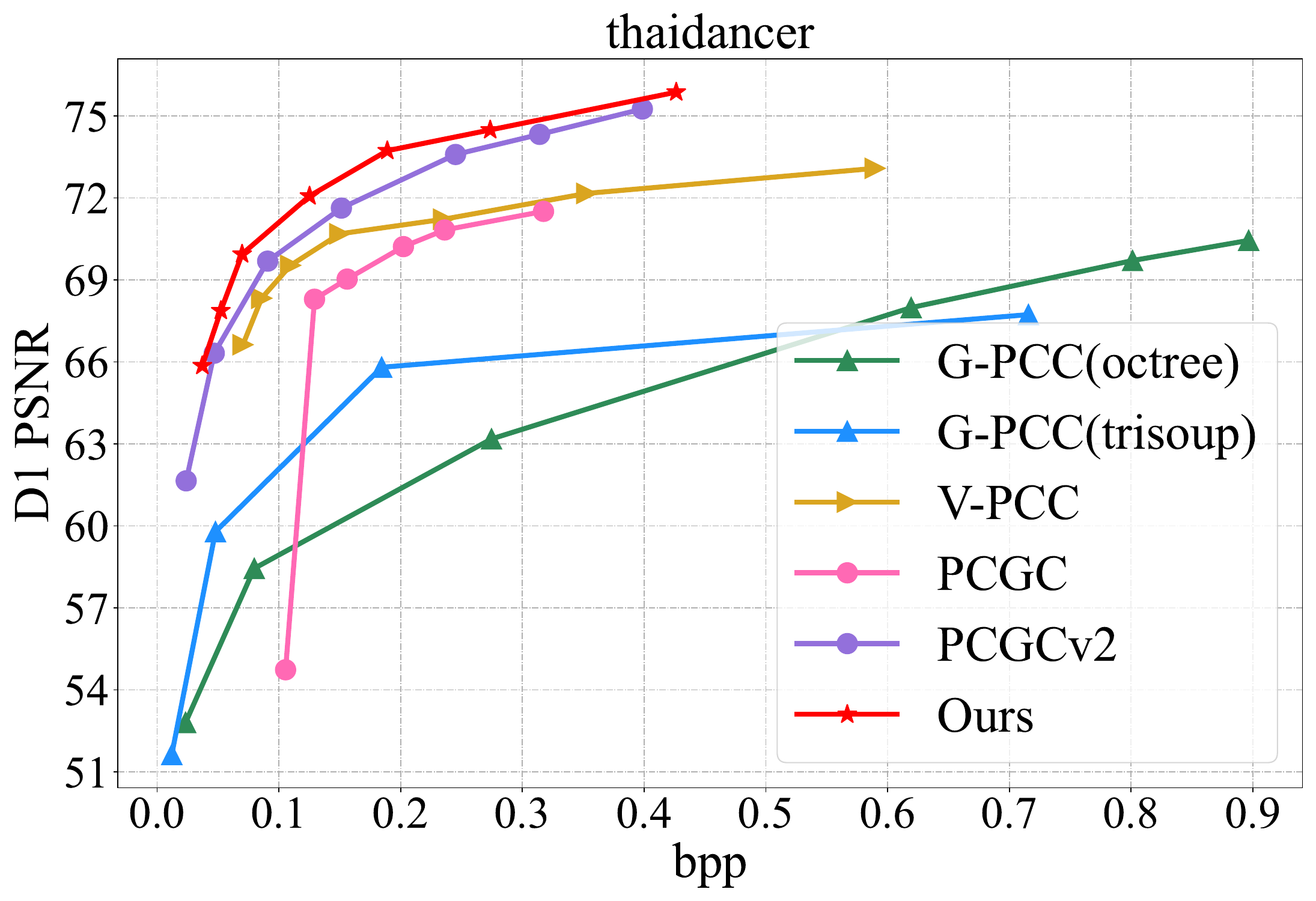}
    \end{minipage}
    \begin{minipage}[]{0.31\linewidth}
        \includegraphics[width=\linewidth]{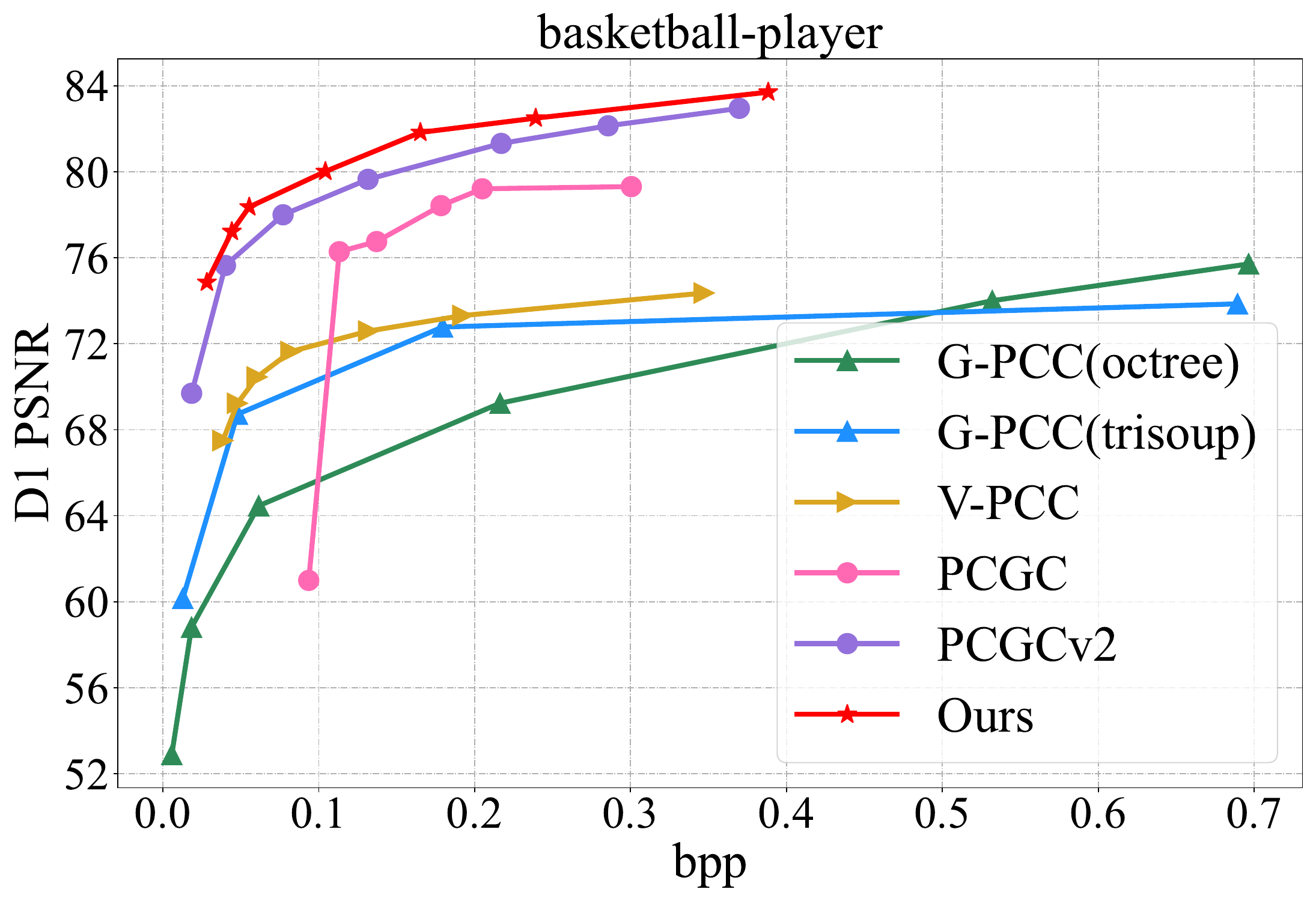}
    \end{minipage}

    \begin{minipage}[]{0.31\linewidth}
        \includegraphics[width=\linewidth]{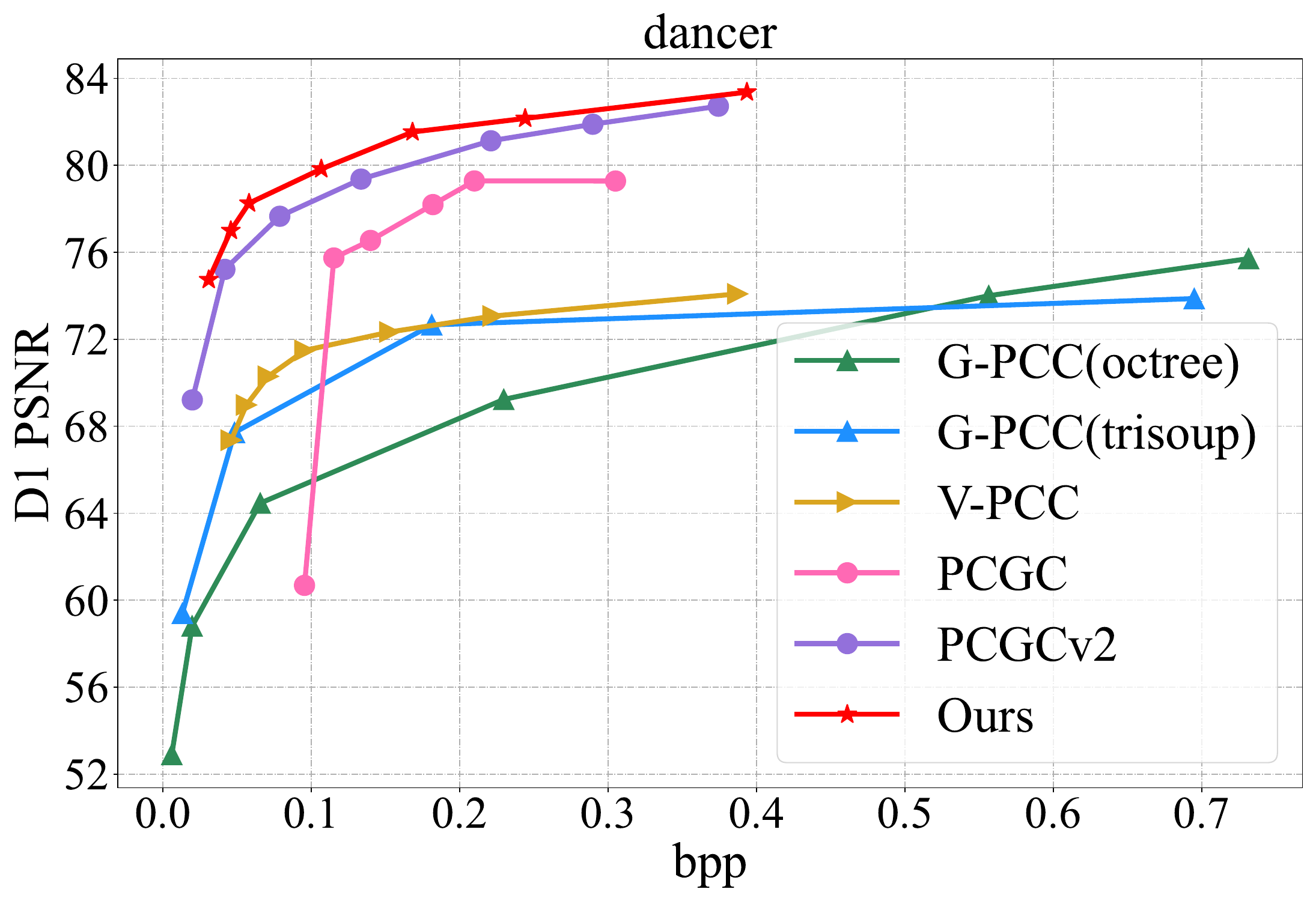}
    \end{minipage}
    \begin{minipage}[]{0.31\linewidth}
        \includegraphics[width=\linewidth]{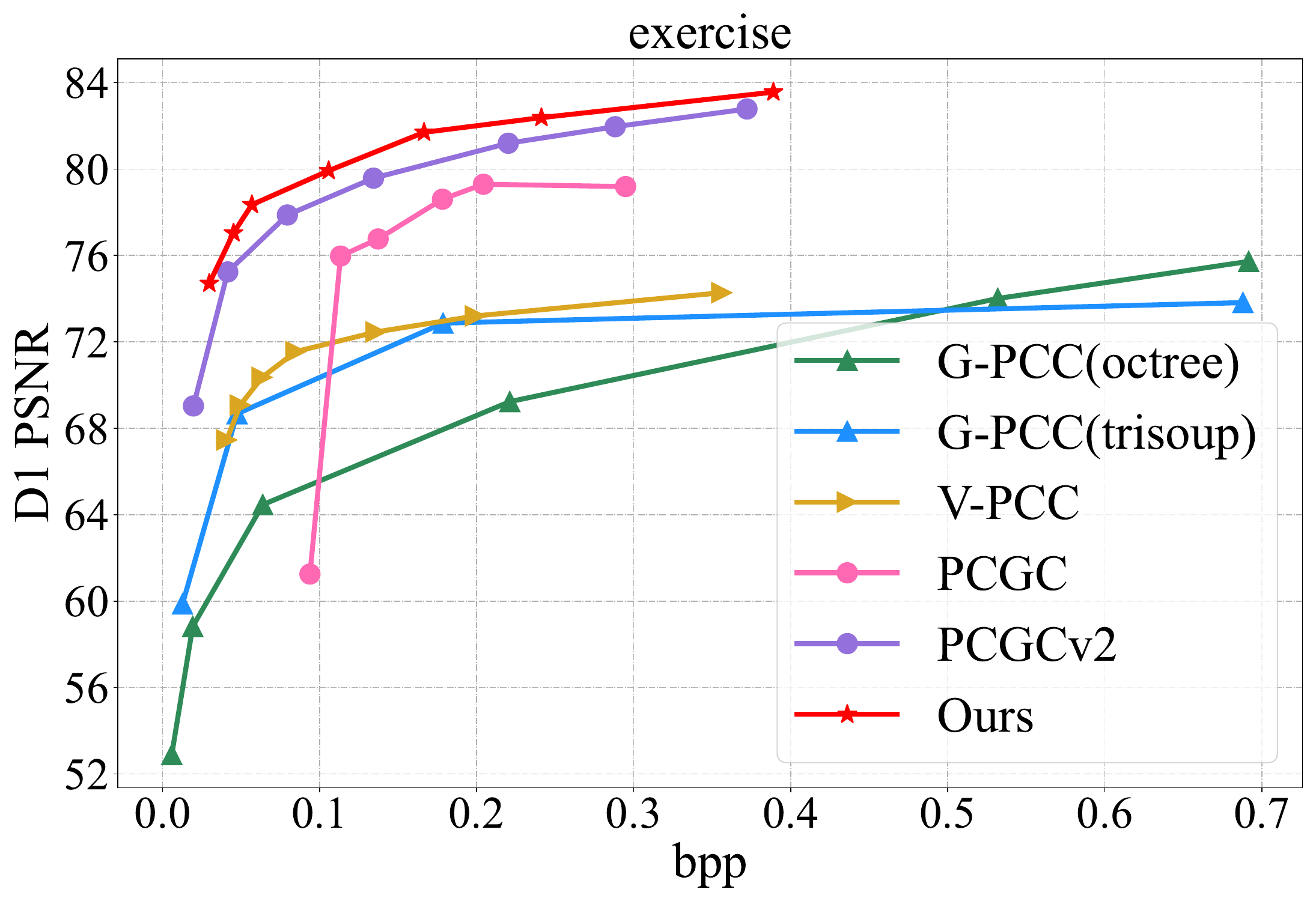}
    \end{minipage} 
    \begin{minipage}[]{0.31\linewidth}
        \includegraphics[width=\linewidth]{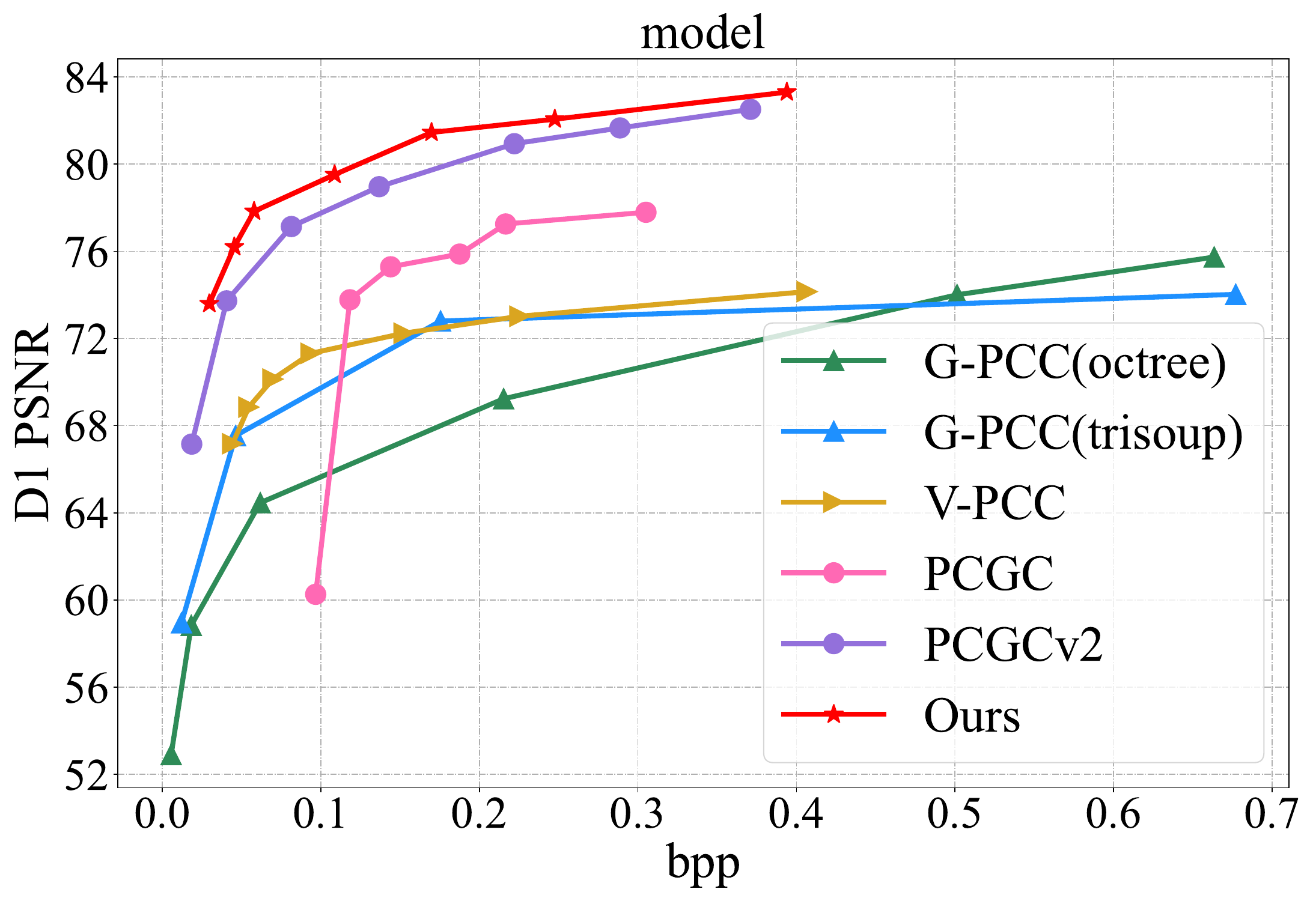}
    \end{minipage}

    % \begin{minipage}[]{0.31\linewidth}
    %     \includegraphics[width=\linewidth]{figures/0100_bpp_D1.png}
    % \end{minipage}
    \begin{minipage}[]{0.31\linewidth}
        \includegraphics[width=\linewidth]{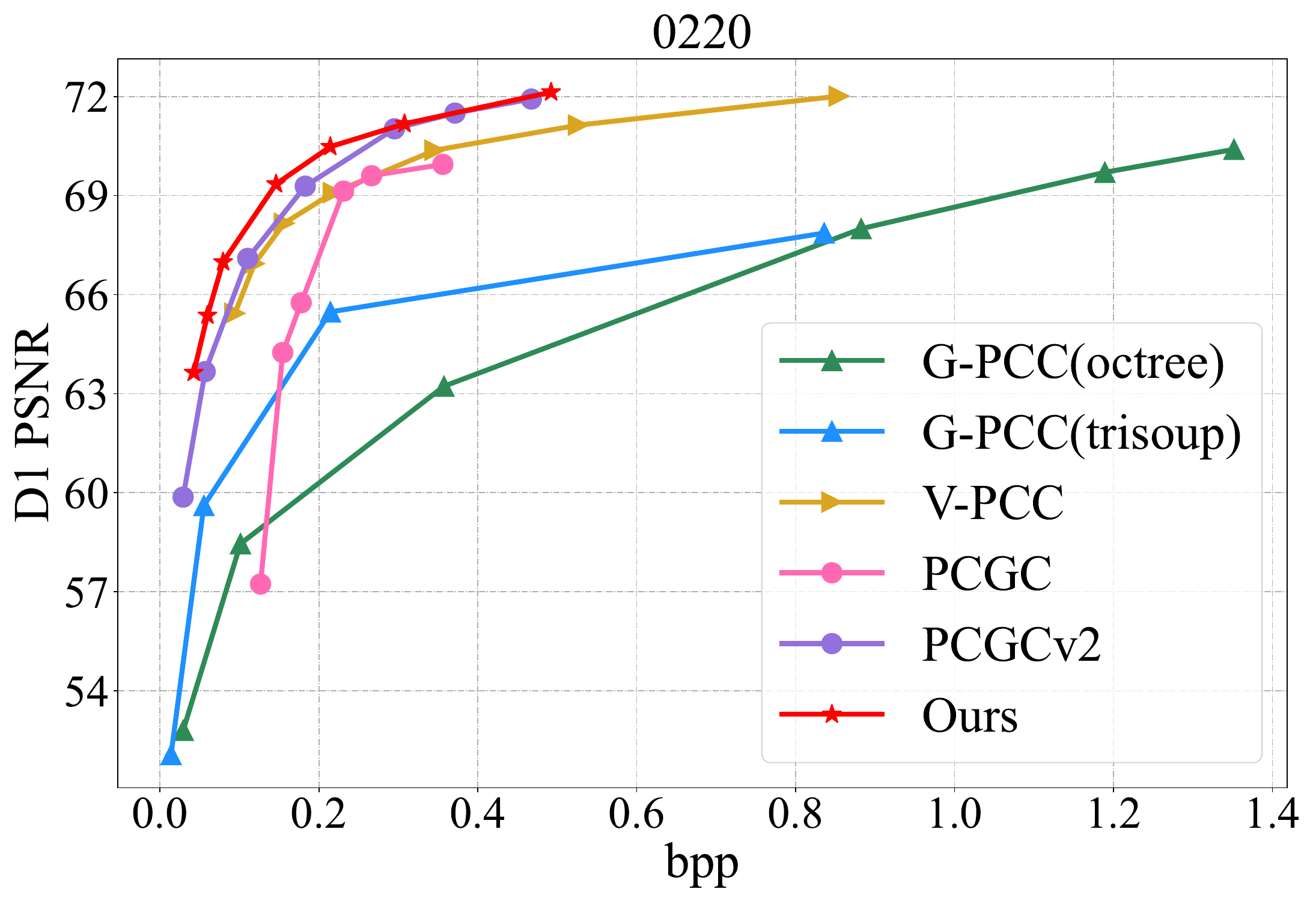}
    \end{minipage}
    \begin{minipage}[]{0.31\linewidth}
        \includegraphics[width=\linewidth]{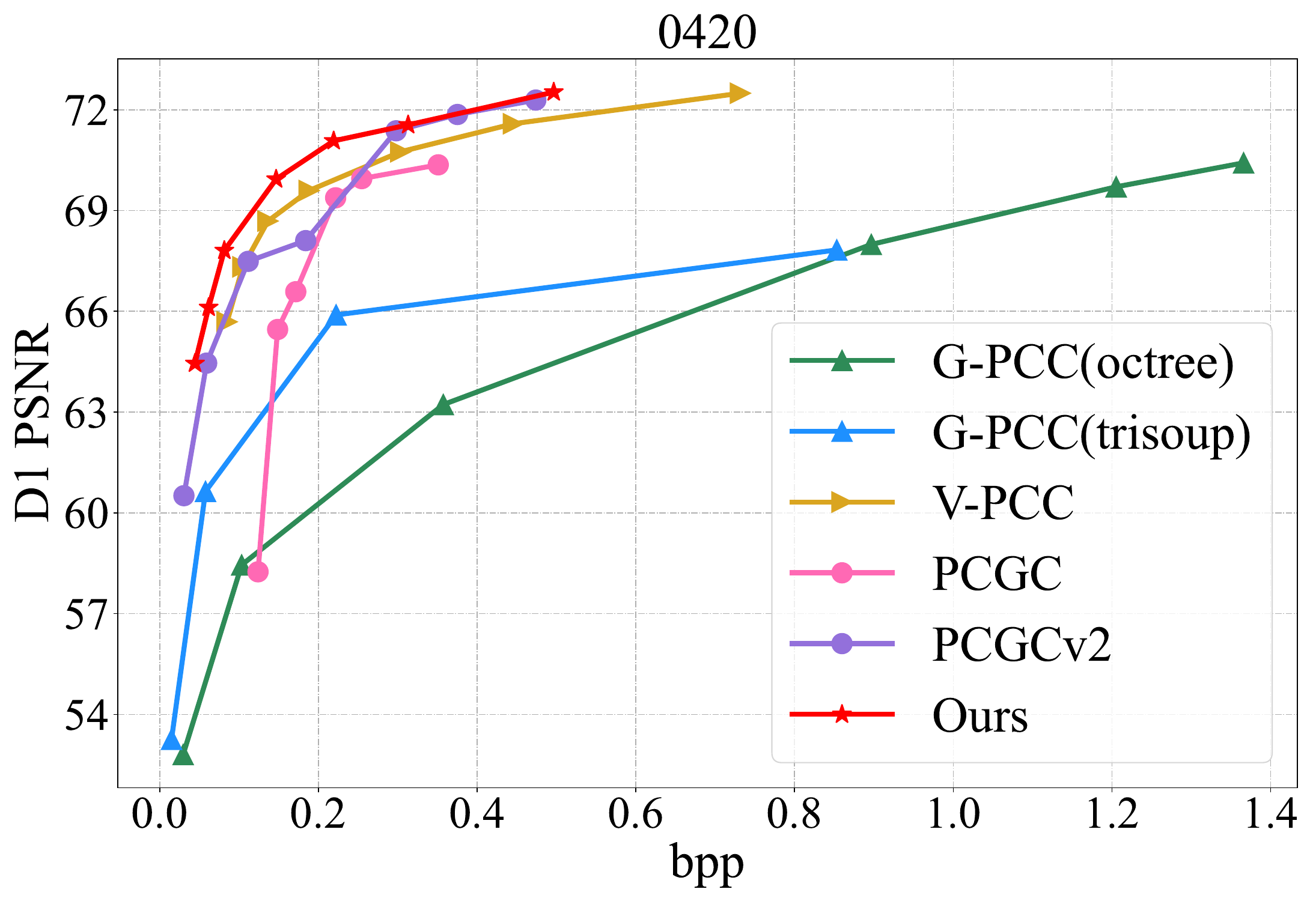}
    \end{minipage}
    \begin{minipage}[]{0.31\linewidth}
        \includegraphics[width=\linewidth]{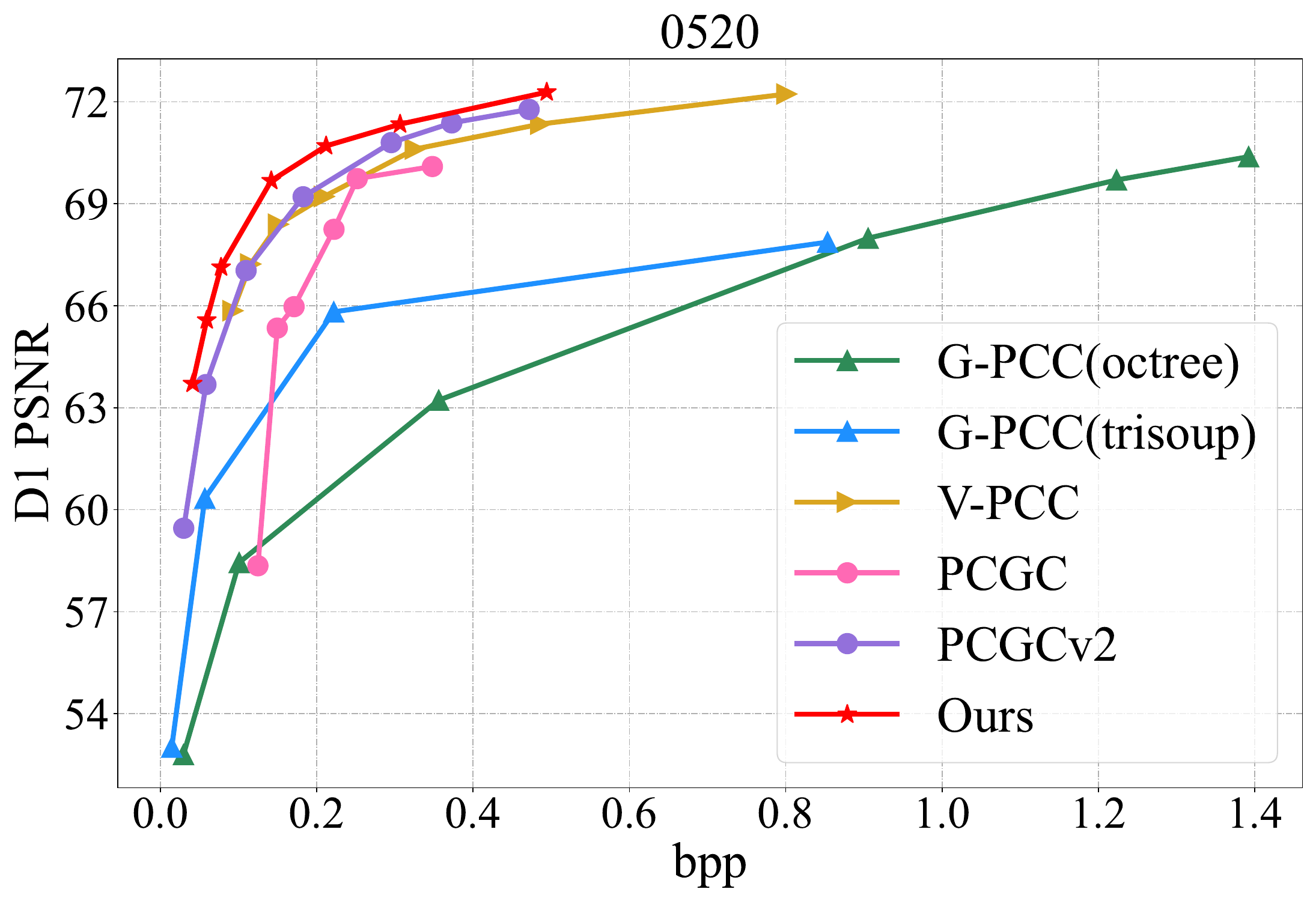}
    \end{minipage}

    \caption{The RD performance of the proposed approach and baselines on the Owlii~\cite{owlii}, 8iVSLF~\cite{8ivslf}, and THuman2.0~\cite{tao2021function4d} datasets using D1 error~\cite{ctc_vpcc, ctc_gpcc}.}
    \label{fig:RD_curves}
\end{figure*}

As shown in Fig.~\ref{fig:RD_curves}, our proposed framework yields superior RD performance compared with other traditional and learning-based methods on diverse human point clouds in terms of D1 PSNR.
Furthermore, our approach and PCGCv2 outperform traditional codecs, while PCGC only falls behind V-PCC. This demonstrates the promising capability of learning-based point cloud geometry compression methods, which can represent point clouds as a sparse set of points equipped with learned feature embeddings. Learning-based schemes unlock avenues for efficiently encoding 3D geometry via end-to-end neural networks, especially at higher bitrates where they can better preserve geometric details than conventional codecs.

\begin{figure}[!t]
    \centering
    \includegraphics[width=0.8\linewidth]{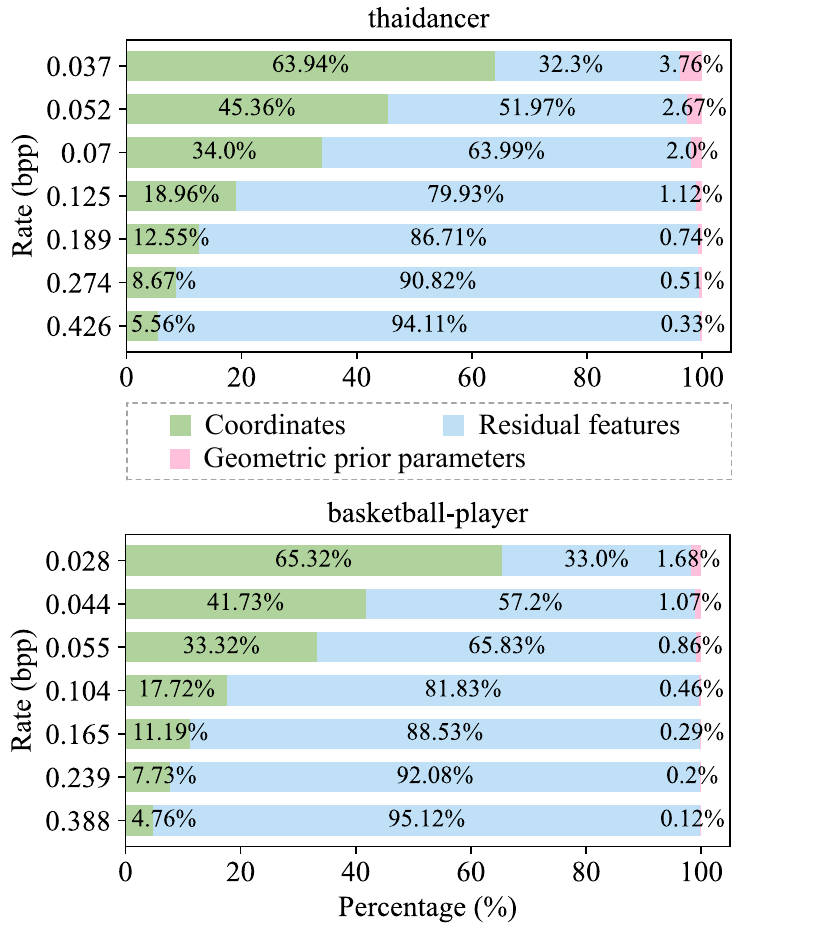}
    \caption{The bitstream composition at different bitrate levels.}
    \label{fig:composition}
\end{figure}

\subsection{Ablation Studies}\label{sec:ablation}

\begin{figure}[!t]
    \centering
    \includegraphics[width=\linewidth]{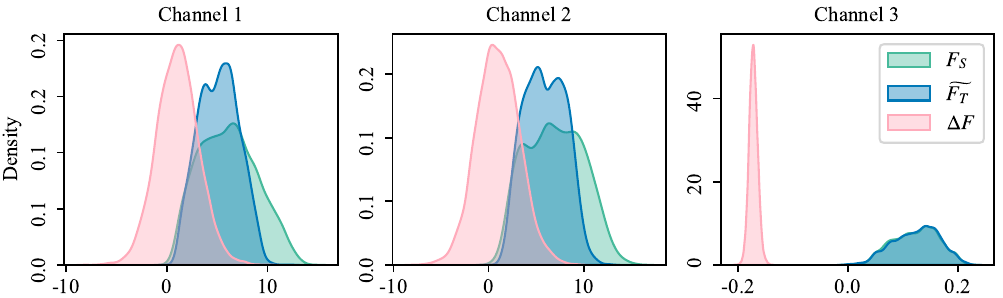}
    \caption{The distributions of features of the source point cloud $\mathbf{F}_\mathbf{S}$, warped features of the aligned point cloud $\widetilde{\mathbf{F}}_\mathbf{T}$, and residual features $\Delta \mathbf{F}$ in different channels.}
    \label{fig:feat_distritbution}
\end{figure}

\begin{figure}[!t]
    \centering
    \includegraphics[width=\linewidth]{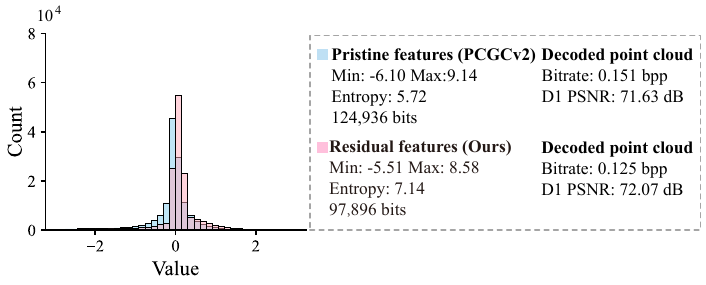}
    \caption{The histogram, value range, entropy, and corresponding decoded point cloud information of pristine features and residual features.}
    \label{fig:features}
\end{figure}

To further validate the effectiveness of our proposed scheme, we provide the bitstream composition, residual features, visualization results, RD performance on point clouds with different geometry precisions, feature channels, runtime comparisons, and performance of animal point clouds.

\begin{figure}[!t]
    \centering
    \includegraphics[width=\linewidth]{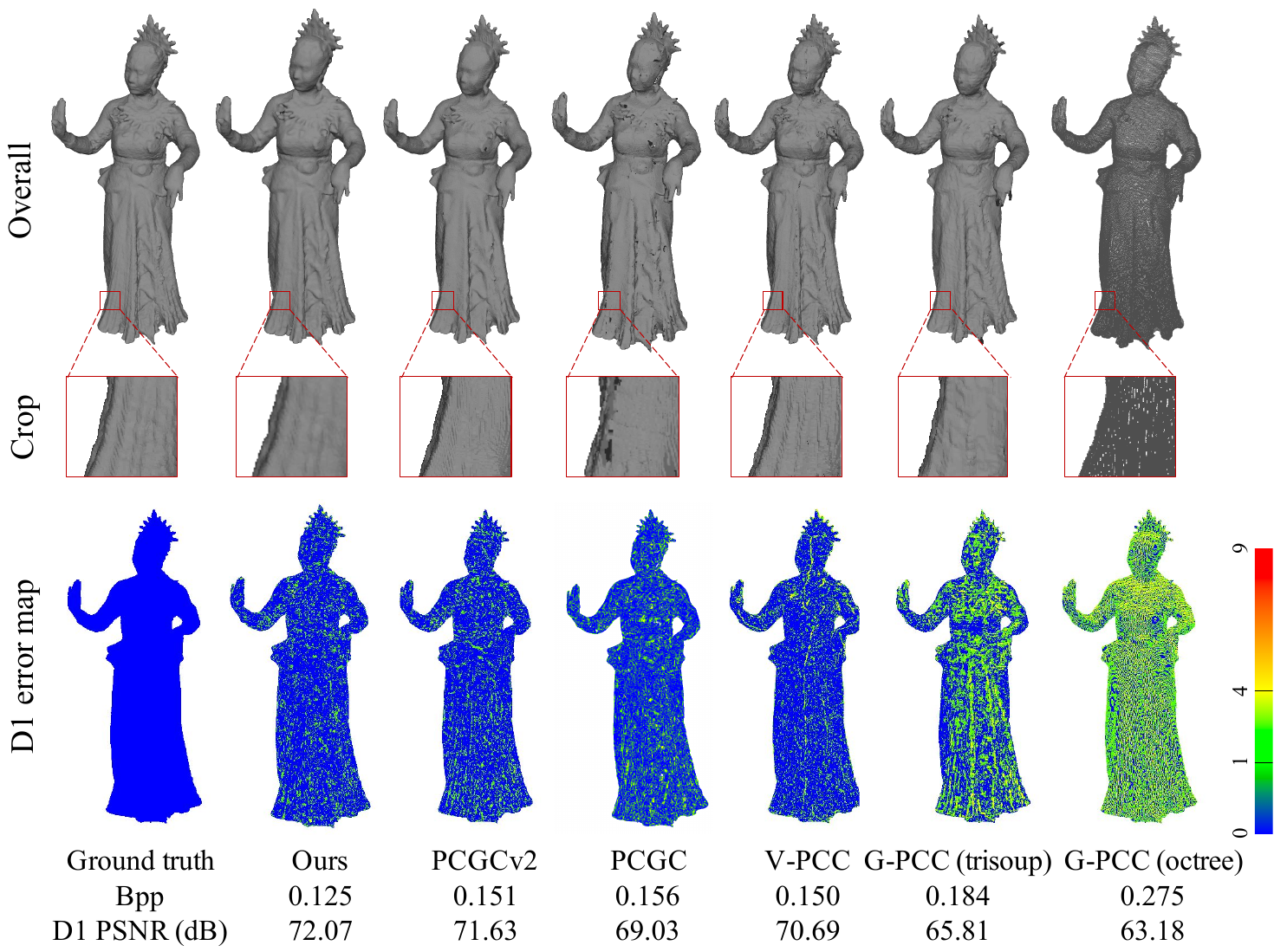}
    \caption{Visualization of geometry reconstruction results of the sequence \textit{thaidancer} from our method, PCGCv2, PCGC, V-PCC, G-PCC (trisoup), and G-PCC (octree). It is worth mentioning that areas within the red rectangles in the first row are magnified in the second row. The final row exhibits error maps between the reconstructed point clouds and ground truth in terms of D1.}
    \label{fig:visualization_thaidancer}
\end{figure}

\begin{figure}[!t]
    \centering
    \includegraphics[width=\linewidth]{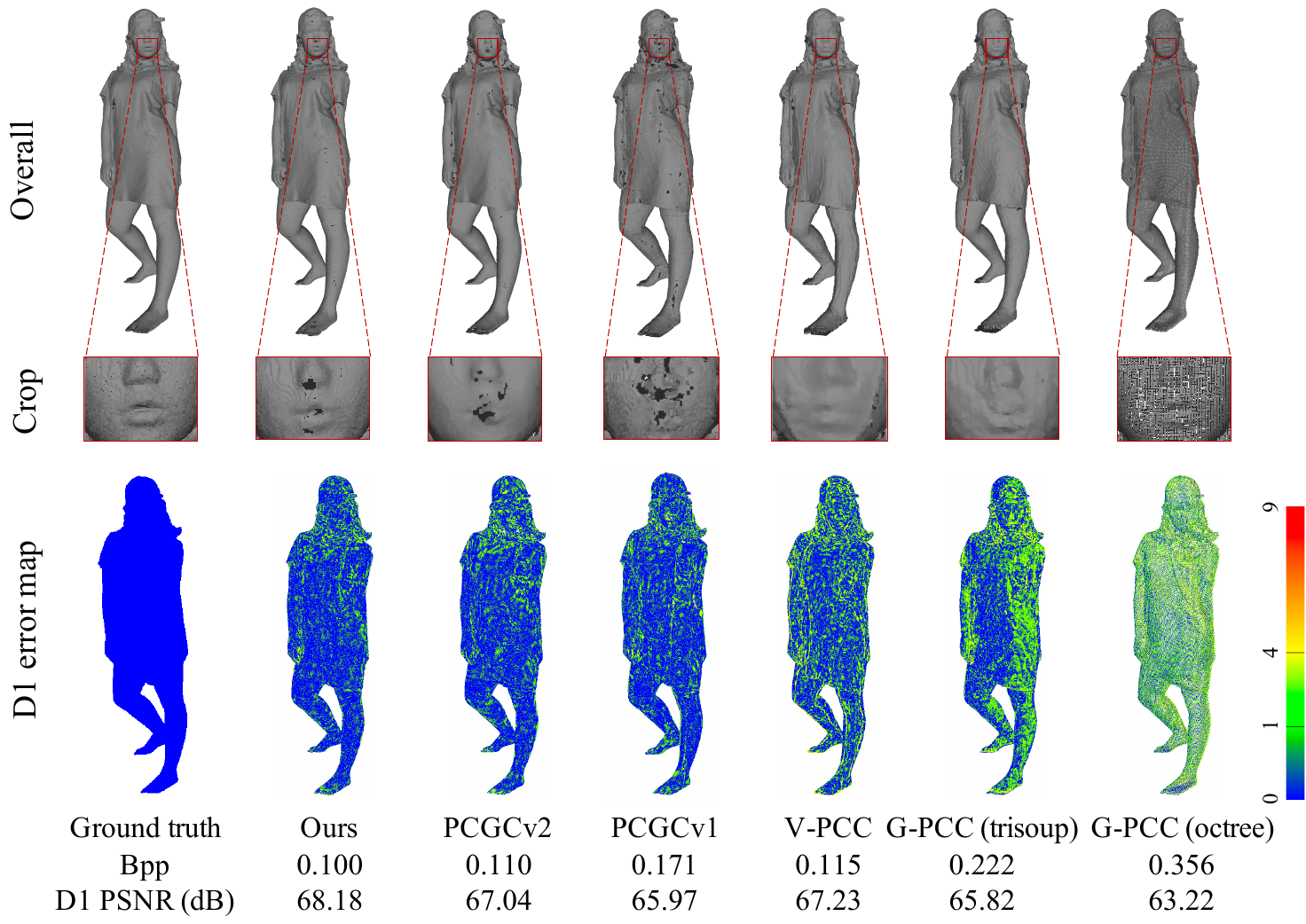}
    \caption{Visualization of geometry reconstruction results of the sequence \textit{0520}. Areas within the red rectangles in the first row are magnified in the second row. The final row exhibits error maps between the reconstructed point clouds and ground truth in terms of D1.}
    \label{fig:visualization_0520}
\end{figure}

\subsubsection{Bitstream composition}

To investigate the cost of geometric priors introduced in our approach, we present the bitstream composition at different bitrate levels, as illustrated in Fig.~\ref{fig:composition}. For each bitrate level, we report the percentage of bits in terms of downsampled coordinates, residual features, and geometric prior parameters. In particular, we observe that geometric parameters account for a small portion of the total bits, with less than 3.8\% in the sequence \textit{thaidancer} and at most 1.7\% in the sequence \textit{basketball-player}. More importantly, it is observed that the proportion of bits allocated to geometric prior parameters decreases as the bitrate increases. This occurs because the quantized 86 parameters require approximately $1,368$ bits, and residual features become the primary consumer of bits. For higher bitrates, bits of geometric prior parameters can take up less than 0.5\%, while residual features occupy more than 90\%. This demonstrates that our method has the potential to reduce the number of bits needed for features with negligible cost by utilizing geometric prior parameters.

\subsubsection{Analysis of residual features}

Fig.~\ref{fig:feat_distritbution} showcases the distribution of two features before and after residual feature computation, as described in Eqn.~(\ref{eq:residual}). The residual feature $\Delta \mathbf{F}$ in our framework, represented by the pink area in Fig.~\ref{fig:feat_distritbution}, has a more concentrated distribution in different channels compared to the feature of the source point cloud $\mathbf{F}_\mathbf{S}$ and the warped feature of the aligned point cloud $\widetilde{\mathbf{F}}_\mathbf{T}$. As the residual features are further encoded by the entropy bottleneck, we compare two cases: compressing pristine features with PCGCv2 and compressing residual features with our approach. The histogram in Fig.~\ref{fig:features} shows that the residual feature has more values near zero and a limited value range. As a result, the entropy of the residual feature is smaller at 14.09 compared to 15.24 for the pristine feature. Furthermore, although the residual feature requires fewer bits at $97,896$ compared to $124,936$ for the pristine feature, the reconstructed point cloud has better quality with 0.44~dB gain in terms of D1 PSNR. This result demonstrates that residual features require fewer bits while maintaining better information fidelity compared to directly compressing pristine features.

\begin{figure}[!t]
    \centering
    %0.45, 0.295
    \begin{minipage}[]{0.48\linewidth}
        \includegraphics[width=\linewidth]{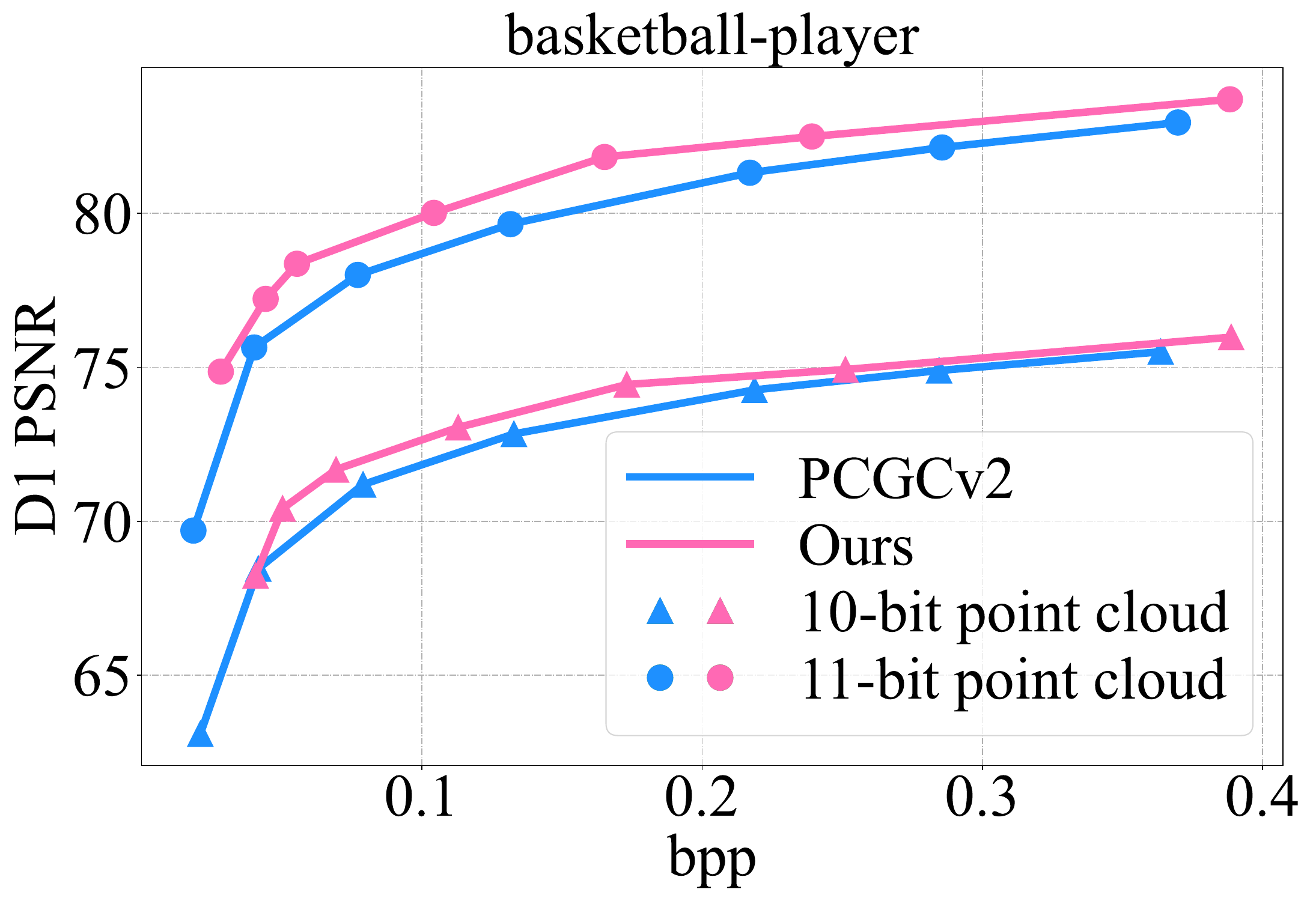}
    \end{minipage}
    \begin{minipage}[]{0.48\linewidth}
        \includegraphics[width=\linewidth]{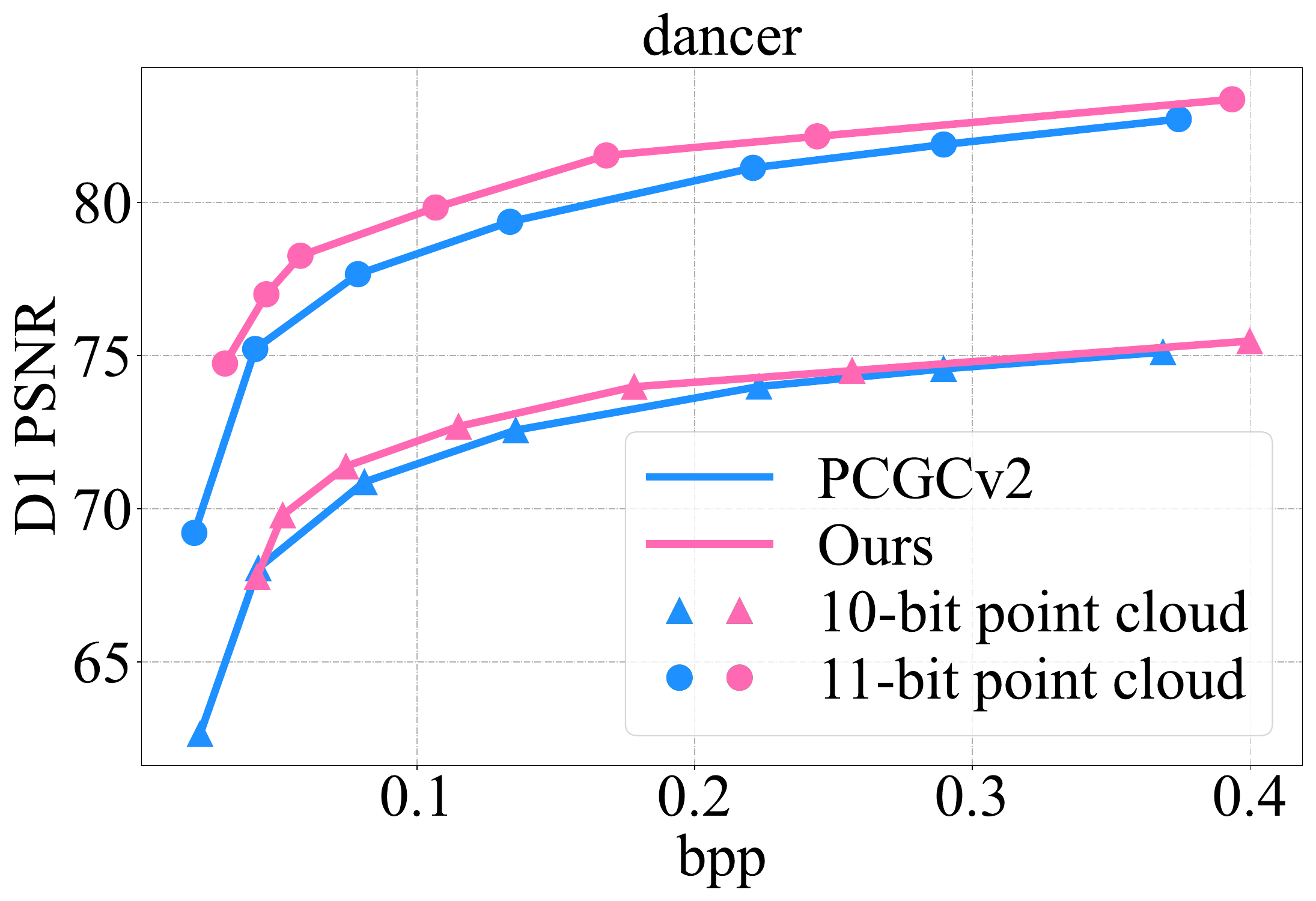}
    \end{minipage}

    \caption{The RD performance of our method and PCGCv2 on sequences with different geometry precision.}
    \label{fig:vox}
\end{figure}

\begin{table}[!t]
    \caption{RD results of the proposed methods with various bottleneck channels against the baseline PCGCv2 on the Owlii, 8iVSLF, and THuman2.0 datasets using D1 and D2 errors}
    \label{tbl:channel16}
    % \small
    \centering
    \renewcommand{\arraystretch}{1.5}
    \resizebox{0.49\textwidth}{!}
    {
    \begin{tabular}{c|c||ccc||ccc}
        \hline \hline
        \multicolumn{2}{c||}{Methods}             & \multicolumn{3}{c||}{Ours (channels=8)} & \multicolumn{3}{c}{Ours (channels=16)}   \\ \hline
        \multicolumn{2}{c||}{Sequences}           & 10-vox  & 11-vox  & \cellcolor[HTML]{f7f7f7}{Average}  & 10-vox   & 11-vox  & \cellcolor[HTML]{f7f7f7}{Average}    \\ \hline
        & \begin{tabular}[c]{@{}c@{}}BD-Rate
            %\\(\%)
        \end{tabular} & {\ul -27.31} & -30.79 & \cellcolor[HTML]{F7F7F7}{\ul -28.70} & -21.46 & {\ul -32.62} & \cellcolor[HTML]{F7F7F7}-25.92  \\
        \multirow{-2}{*}{\begin{tabular}[c]{@{}c@{}}D1\\PSNR\end{tabular}} & \begin{tabular}[c]{@{}c@{}}BD-PSNR
            %\\(dB)
        \end{tabular}  & {\ul 1.20}   & 1.31   & \cellcolor[HTML]{F7F7F7}{\ul 1.25}   & 0.88   & {\ul 1.20}   & \cellcolor[HTML]{F7F7F7}1.01    \\ \hline
                                  & \begin{tabular}[c]{@{}c@{}}BD-Rate
                                    %\\(\%)
                                \end{tabular} & {\ul -19.91} & -26.63 & \cellcolor[HTML]{F7F7F7}{\ul -22.60} & -14.66 & {\ul -31.46} & \cellcolor[HTML]{F7F7F7}-21.38  \\
        \multirow{-2}{*}{\begin{tabular}[c]{@{}c@{}}D2
            %\\PSNR
        \end{tabular}} & \begin{tabular}[c]{@{}c@{}}BD-PSNR
            %\\(dB)
        \end{tabular}  & {\ul 0.97}   & 1.35   & \cellcolor[HTML]{F7F7F7}{\ul 1.12}   & 0.67   & {\ul 1.35}   & \cellcolor[HTML]{F7F7F7}0.94   
        \\
        \hline \hline   
    \end{tabular}
    }
\end{table}

\begin{table}[!t]
\caption{The average running time (s) in different approaches}
\centering
\resizebox{\linewidth}{!}
{
\renewcommand{\arraystretch}{1.25}
% \footnotesize
\begin{tabular}{c||c c c c c c}
\hline
\hline
          & \begin{tabular}[c]{@{}c@{}}G-PCC\\(octree)\end{tabular} & \begin{tabular}[c]{@{}c@{}}G-PCC\\(trisoup)\end{tabular}& V-PCC & PCGC & PCGCv2 & Ours \\ \hline
Enc. &      3.15	          &       16.1          &  82.63    & 32.30 &  1.5       & 12.4     \\ \hline
% Enc. &      3.15	          &       16.10          &  82.63     &  1.50      &  13.30     \\ \hline
Dec. &      1.1          &       13.21          &  2.09     & 17.37  &  0.77   &  2.76    \\ \hline
\hline
\end{tabular}
}
\label{tbl:time}
\end{table}

\subsubsection{Qualitative evaluations}
We visualize the reconstructed point clouds from different point cloud geometry compression methods. Fig.~\ref{fig:visualization_thaidancer} and Fig.~\ref{fig:visualization_0520} display the overall geometry of the whole point cloud, a zoomed-in region with geometry details, and an error map in terms of D1 distance for the sequences \textit{thaidancer} and \textit{0520}, respectively. Compared to other baselines, our proposed approach can generate high-quality decoded point cloud geometry with lower bitrates. Fig.~\ref{fig:visualization_thaidancer} shows that our method better reconstructs the pleats on the skirt with the least bpp, while the same regions are smoother with PCGCv2 and visible holes are introduced with PCGCv1. Although V-PCC achieves satisfactory reconstruction results in local regions at a higher bitrate, there are apparent cracks in the vertical middle due to the patch generation operations. While G-PCC (octree) leads to a massive loss of points, G-PCC (trisoup) yields comparable visualization results overall but produces cluttered protruding local areas. The visualization of the sequence \textit{0520} also shows similar results in Fig.~\ref{fig:visualization_0520}. For instance, our proposed method reconstructs the clear shape of the nose and mouth, while the results from PCGCv2 are smoother with more holes. Additionally, there are obvious distortions on 3D block boundaries from PCGCv1, as this approach depends on the cube partition of a point cloud during inference.

\begin{table*}[hbt]
    \caption{BD-Rate and BD-PSNR results on animal point clouds against the baselines G-PCC (octree)~\cite{gpcc}, G-PCC (trisoup)~\cite{gpcc}, V-PCC~\cite{vpcc}, PCGC~\cite{wang2021lossy}, and PCGCv2~\cite{wang2021multiscale} using D1 error~\cite{ctc_vpcc, ctc_gpcc}}
    \centering
    % \resizebox{\textwidth}{!}
    { % \linewidth
\renewcommand{\arraystretch}{1.15}
\footnotesize
    \begin{tabular}{c||ccccc||ccccc}
    \hline
\hline

\multirow{2}{*}{Sequence}    & \multicolumn{5}{c||}{BD-Rate with D1 PSNR (\%)}       & \multicolumn{5}{c}{BD-PSNR with D1 (dB)}    \\ \cline{2-11} 
    & \begin{tabular}[c]{@{}c@{}}G-PCC\\(octree)\end{tabular} & \begin{tabular}[c]{@{}c@{}}G-PCC\\(trisoup)\end{tabular} & V-PCC           & PCGC            & PCGCv2          & \begin{tabular}[c]{@{}c@{}}G-PCC\\(octree)\end{tabular} & \begin{tabular}[c]{@{}c@{}}G-PCC\\(trisoup)\end{tabular} & V-PCC         & PCGC          & PCGCv2        \\ \hline

Dog                       & -92.60          & -91.47          & -59.25          & -63.08          & -7.19           & 12.22         & 7.99           & 4.53          & 5.06          & 0.28          \\
Cow                       & -90.16          & -73.96          & -17.46          & -57.90          & -17.81          & 9.30          & 5.14           & 0.48          & 2.49          & 0.73          \\
Horse                     & -78.61          & -6.71           & -13.57          & -48.50          & -12.78          & 5.74          & 0.60           & 0.31          & 1.33          & 0.61          \\ \hline
\textbf{Average with D1}  & \textbf{-87.12} & \textbf{-57.38} & \textbf{-30.09} & \textbf{-56.49} & \textbf{-12.59} & \textbf{9.09} & \textbf{4.58}  & \textbf{1.77} & \textbf{2.96} & \textbf{0.54} \\ 

\hline
\hline
\end{tabular}

    }
    \label{tbl:RD_animal}
\end{table*}

\begin{figure*}[htb]
    \centering
    %0.45, 0.31
    \begin{minipage}[]{0.31\linewidth}
        \includegraphics[width=\linewidth]{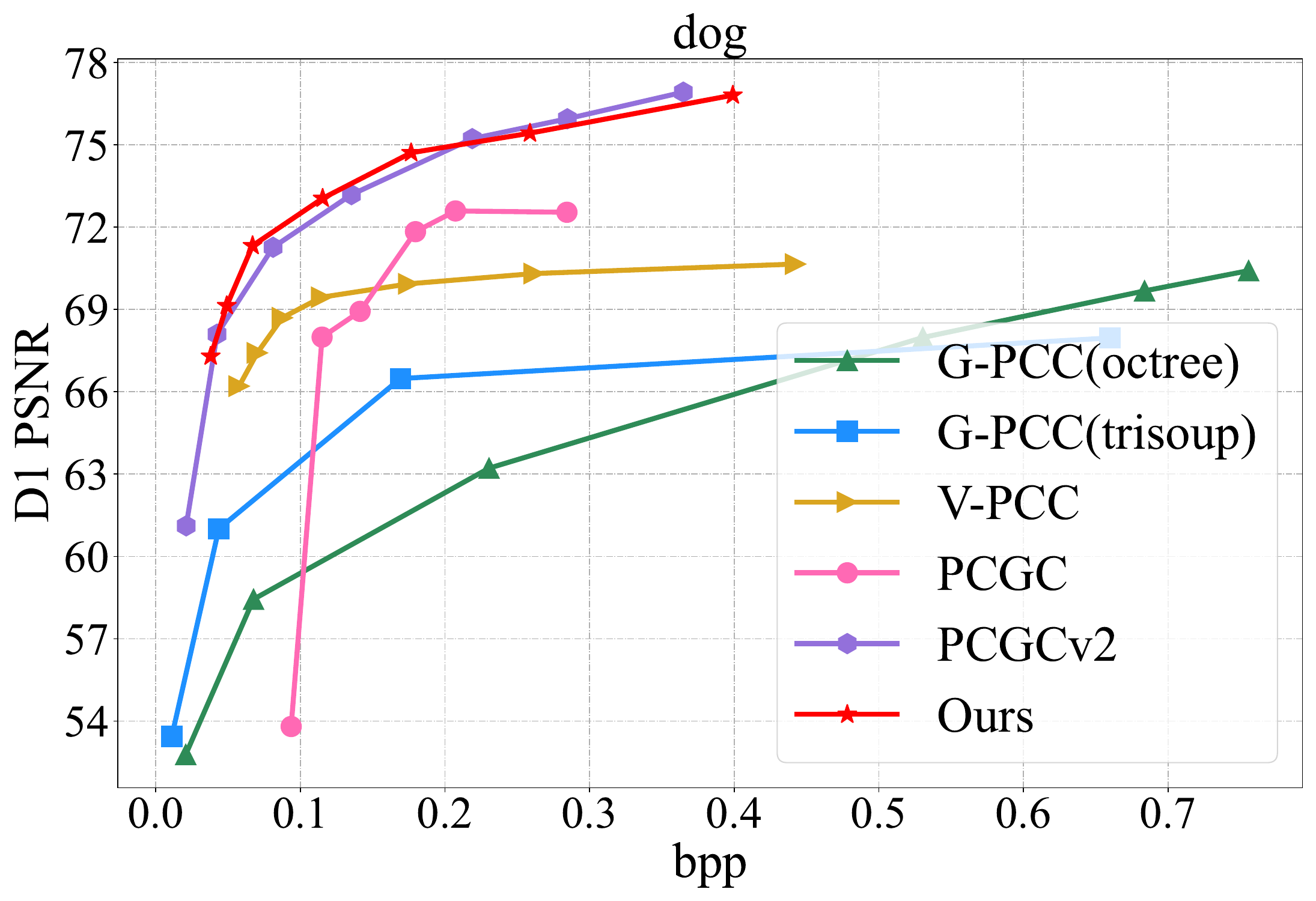}
    \end{minipage}
    \begin{minipage}[]{0.31\linewidth}
        \includegraphics[width=\linewidth]{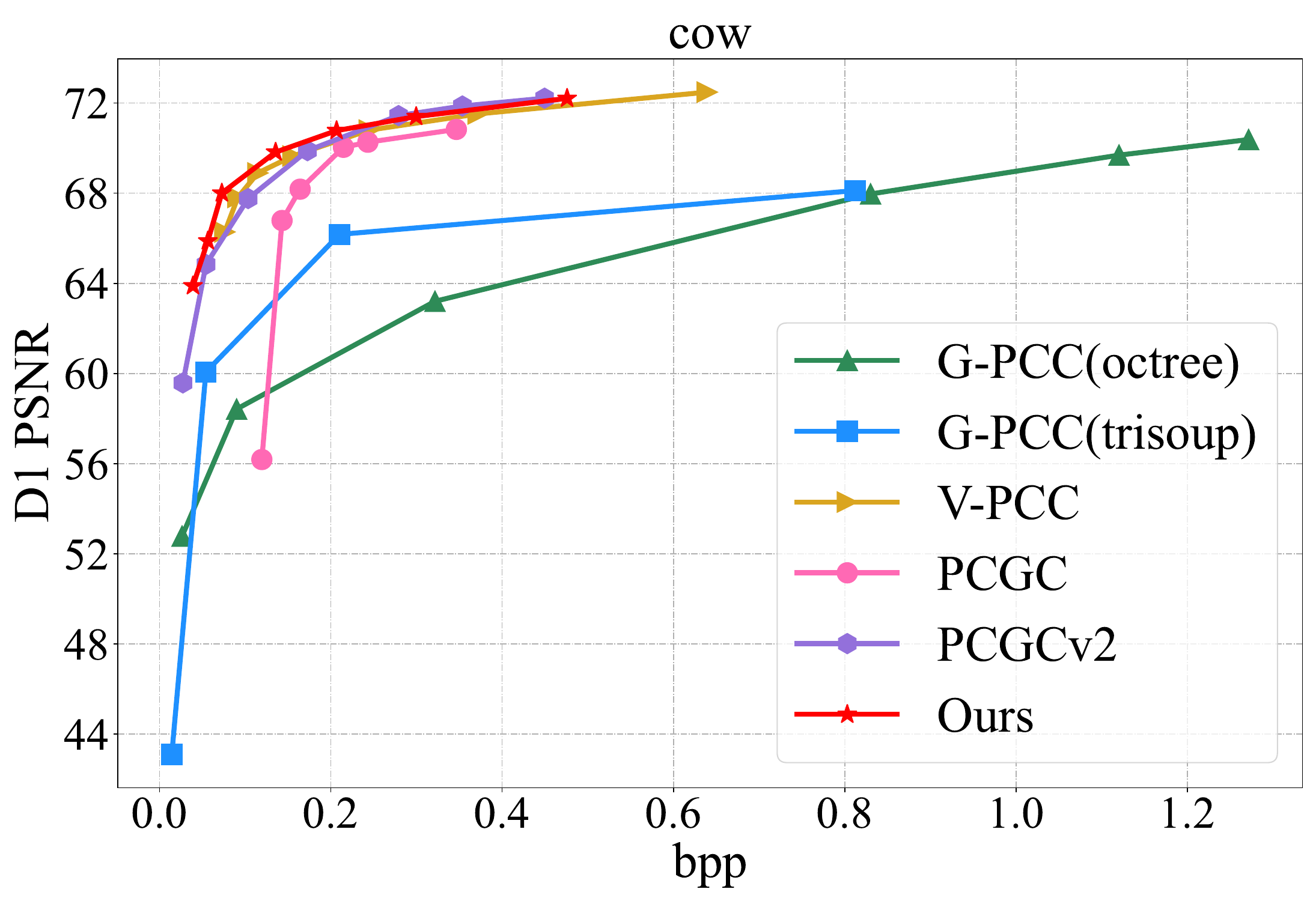}
    \end{minipage}
    \begin{minipage}[]{0.31\linewidth}
        \includegraphics[width=\linewidth]{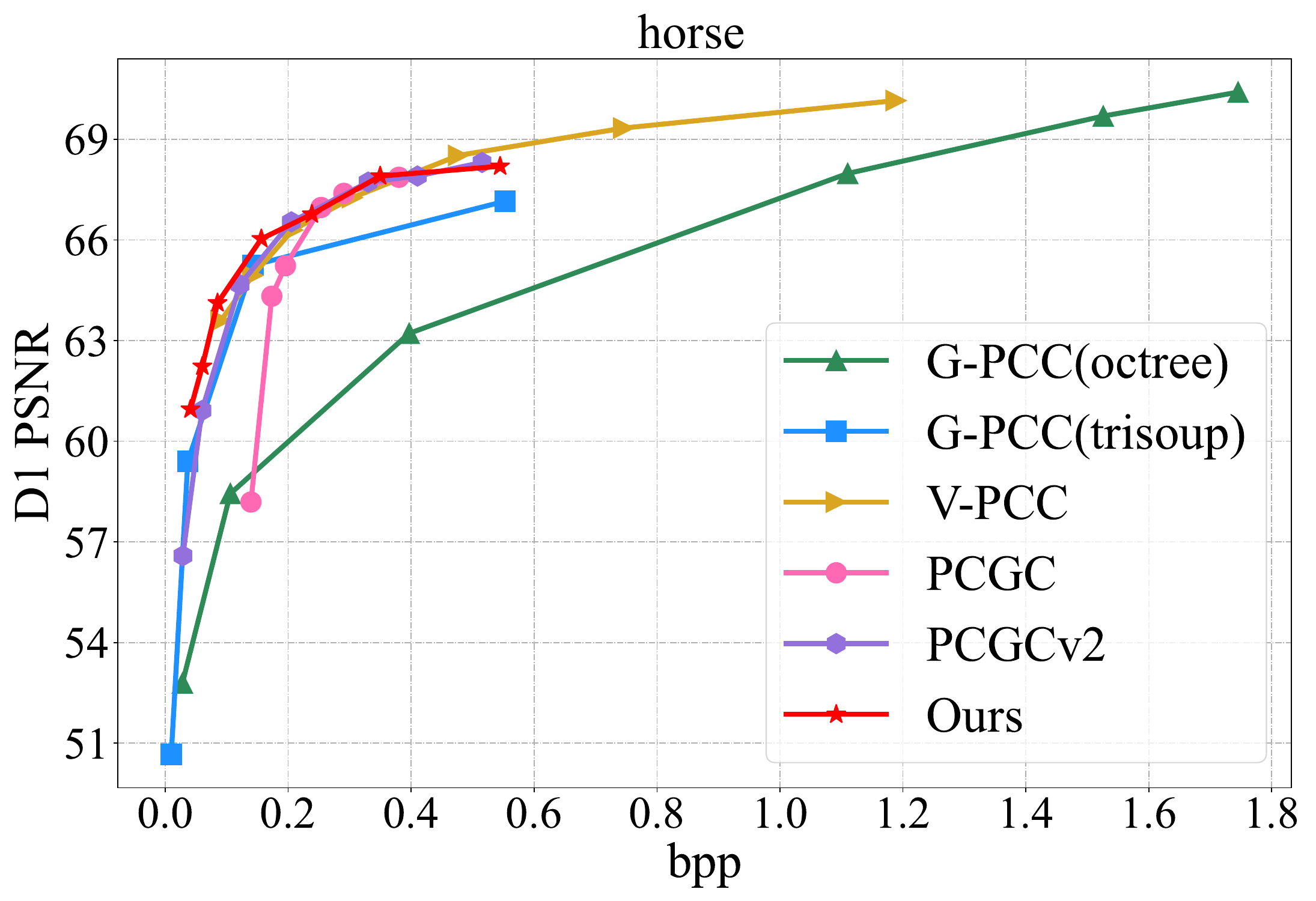}
    \end{minipage}
    \begin{minipage}[]{0.31\linewidth}
        \includegraphics[width=\linewidth]{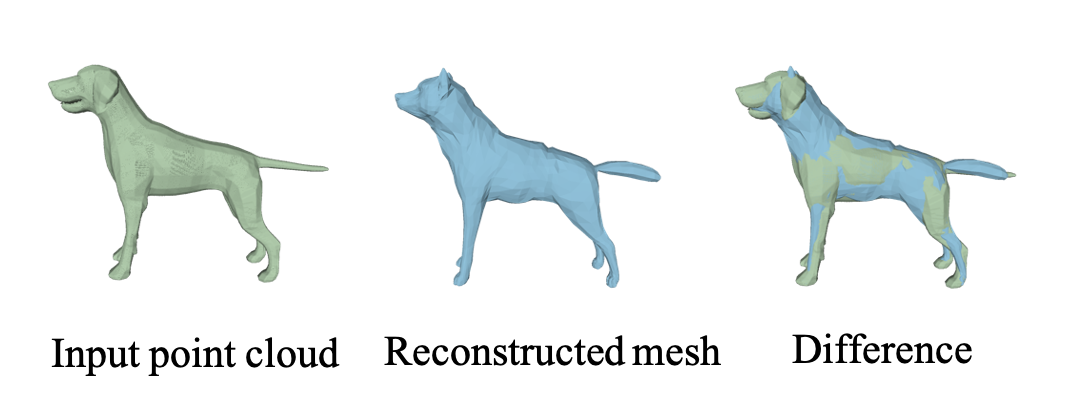}
    \end{minipage}
    \begin{minipage}[]{0.31\linewidth}
        \includegraphics[width=\linewidth]{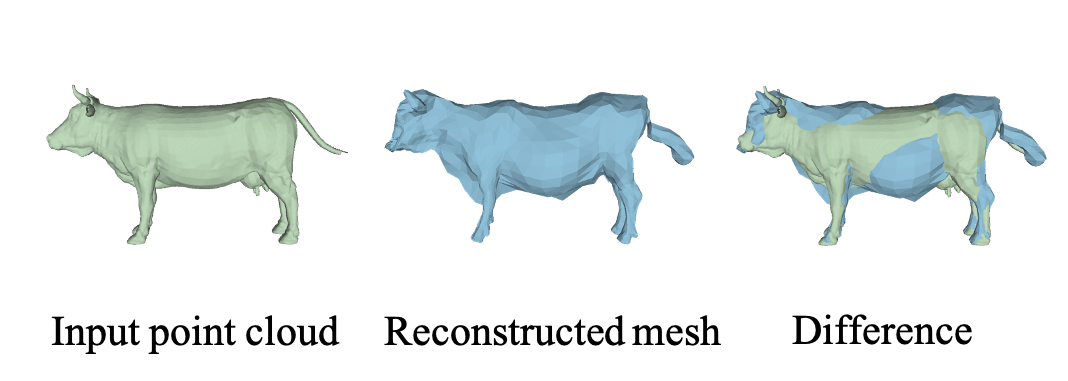}
    \end{minipage}
    \begin{minipage}[]{0.31\linewidth}
        \includegraphics[width=\linewidth]{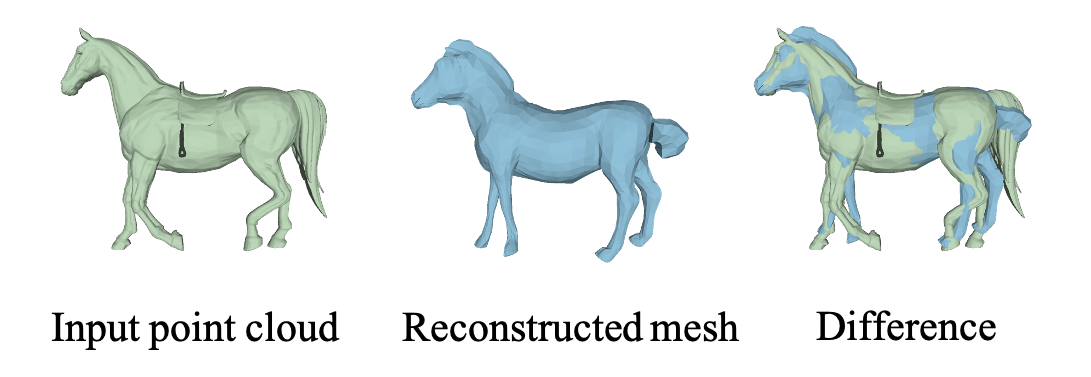}
    \end{minipage}
    \caption{The RD performance of the proposed approach and baselines on three animal point clouds using D1 error~\cite{ctc_vpcc, ctc_gpcc}.}
    \label{fig:RD_curves_animal}
\end{figure*}

\subsubsection{Geometry precision}
To further investigate the effectiveness of the proposed method, we also compare the performance to PCGCv2 using point clouds of the same sequences with different geometry precision levels. As shown in Fig.~\ref{fig:vox}, our proposed scheme demonstrates improved coding performance for both 10-bit and 11-bit point clouds. More specifically, our scheme achieves 29.31\% bit savings for an 11-bit point cloud \textit{basketball player} and 19.08\% for its 10-bit version compared to PCGCv2. This is because higher geometry precision booms the amount of data needed to compress point clouds. These results are consistent with those shown in Table~\ref{tbl:RD}, where BD-Rate gains for the Owlii dataset are much higher than for other datasets. As human point clouds with higher geometry precision allow for larger and finer granularity of 3D coordinates, our method facilitates the reconstruction of high-accuracy human point clouds.

\subsubsection{Feature channels}
To evaluate the impact of compressing different feature lengths, we conduct an experiment using 16 channels in the bottleneck layer of the encoder, as channels of this layer directly determine the number of elements for compression.
As evidenced in Table~\ref{tbl:channel16}, the 16-channel model performs comparably to our 8-channel model, with both achieving over 20\% BD-Rate gains versus PCGCv2. The additional channels help represent intricate local geometry, especially benefiting higher-precision point clouds. Specifically, our method with 8 channels shows slightly better performance on the 10-vox sequences, while ours with 16 channels is better for 11-vox sequences, as shown in Table~\ref{tbl:channel16}.

\subsubsection{Runtime comparisons}
We further compare the running time of our proposed method and other baseline approaches. We conduct the experiments on a server with an Intel Core i7-10700 CPU and an NVIDIA GeForce RTX 3090 GPU. Following~\cite{wang2021multiscale,akhtar2022interframe}, we compute the encoding and decoding time of all testing point clouds at the highest bitrate level since the runtime of G-PCC varies at different bitrate levels. The traditional codecs G-PCC and V-PCC are applied using C++ with a CPU, while learning-based PCGCv2 and our method are implemented using Python with a GPU. As a general indication of computational complexity, Table~\ref{tbl:time} shows that our method increases encoding and decoding time compared to PCGCv2. This occurs because our approach needs to perform additional mesh regression, mesh manipulation, mesh-to-point-cloud conversion, feature extraction, and feature warping in the encoder, and extra mesh manipulation and feature warping are executed in the decoder. The mesh regression and mesh-to-point-cloud conversion methods used are time-consuming, taking approximately 9.7~s and 1.9~s, respectively. Our approach can be further sped up with efficient mesh processing algorithms. Furthermore, it is worth mentioning that G-PCC (trisoup) is also based on surface sampling, and its encoding time (16.1~s) and decoding time (13.21~s) are higher than our method’s encoding time (12.4~s) and decoding time (2.76~s).

\subsubsection{Source point clouds representing animals}
To further evaluate the capability of our framework in modeling other categories beyond humans, we conduct additional experiments using animal point clouds rather than human point clouds. Parametric deformable models for animals have gained research attention in prior arts due to the challenges posed by animals' non-rigid bodies. The skinned multi-animal linear model (SMAL)~\cite{zuffi20173d} provides a parametric deformation representation for the animal models, with parameters succinctly represented as
\begin{equation}
    \bm{\Sigma} = \{\bm{\alpha}, \bm{\beta}, \bm{\theta}, \bm{\delta}, \phi \},
\label{eq:smal}
\end{equation}
where $\bm{\alpha} \in \mathbb{R}^{34 \times 3}$, $\bm{\beta} \in \mathbb{R}^{27}$, $\bm{\theta} \in \mathbb{R}^{3 \times 1}$, $\bm{\delta} \in \mathbb{R}^{3 \times 1}$, $\phi \in \mathbb{R}$ represent joint rotation, shape, global rotation, translation, and family shape, respectively. In total, the animal parametric model requires 136 parameters, 50 more than the 86 human parameters. Unlike humans, animals are less cooperative subjects for 3D scans, resulting in fewer high-quality animal point clouds in current datasets. Moreover, point-to-mesh fitting for animals remains difficult given their natural behaviors, movements, and highly deformable and non-rigid tails. Consequently, using high-quality meshes from Done3D, we fit SMAL parameters through mesh-to-mesh fitting due to the unavailable point-to-mesh reconstruction. Source animal point clouds are generated by subdividing source meshes.

As evidenced in Table~\ref{tbl:RD_animal}, our method outperforms G-PCC (octree), G-PCC (trisoup), V-PCC, PCGC, and PCGCv2 when measured by D1 error as distortion and bpp as bitrate. Notably, our method achieves 87.12\% bitrate gains and 9.09~dB higher PSNR versus G-PCC (octree), representing a major improvement in compression performance. When compared to PCGCv2, our approach reduces bitrate by 12.59\% while improving PSNR by 0.54 dB. These gains validate the effectiveness of our proposed compression framework for both accurately capturing geometries and compactly encoding their 3D structures. Fig.~\ref{fig:RD_curves_animal} presents RD curves and visualizations of source point clouds, reconstructed meshes from decoded parameters, and their differences. 
While the RD performance gain on animal point clouds is less substantial than that on human point clouds, we propose that this result stems from several complicating factors beyond geometric similarity. Key factors include differences in total point counts, pristine quality of source point clouds, variations in point density, and potentially less structured animal shapes. Overall, these quantitative results demonstrate the superiority of our framework over current compression methods for both human and animal point clouds.

\section{Conclusions}
In this work, we propose a novel deep human point cloud geometry compression scheme based on geometric priors. The novelty of our approach lies in representing human point clouds as a combination of geometric priors and structure variations. By using geometric prior parameters that are quite compact, our method is able to perform feature-level residual operations to remove geometric redundancy. The superior RD performance of our scheme is demonstrated by comparing it to traditional and learning-based methods for analyzing human point clouds from various datasets. Our scheme significantly reduces the bitrate while preserving the same level of data quality. The proposed scheme also achieves an improvement in visual quality with finer geometry details in local areas with the same bitrate. A promising area for future work is investigating more advanced techniques to embed shape and pose information to remove geometric redundancies when compressing point clouds beyond humans.

\bibliographystyle{IEEEtran}
\bibliography{reference/manuscript}

\end{document}